\documentclass{elsart5p}
\usepackage{amssymb,amsmath}
\usepackage{boxedminipage}
\usepackage{graphicx}
\usepackage[latin1]{inputenc}
\usepackage{url}

\begin{document}
\begin{frontmatter} 
\title{Off-line studies of the laser ionization of yttrium at the IGISOL facility}
\author[jyfl]{T. Kessler\corauthref{corauth}}, 
\corauth[corauth]{Corresponding author. Tel.: +358-14-2602440; \\ fax: +358-14-2602351.}
\ead{thomas.kessler@phys.jyu.fi}
\author[jyfl]{I.D. Moore},
\author[leuven]{Y. Kudryavtsev},
\author[jyfl,stuk]{K. Peräjärvi}, 
\author[petersburg]{A. Popov},
\author[jyfl]{P. Ronkanen}, 
\author[jyfl]{T. Sonoda\thanksref{presadd}},
\author[manc]{B. Tordoff},
\author[mainz]{K.D.A. Wendt} and
\author[jyfl]{J. Äystö}
\thanks[presadd]{Present address: Instituut voor Kern- en Stralingsfysika, University of Leuven, Celestijnenlaan 200 D, B-3001 Leuven, %%@
Belgium}
\address[jyfl]{Department of Physics, University of Jyväskylä, P.O. Box 35 (YFL), FIN-40014, Finland}
\address[leuven]{Instituut voor Kern- en Stralingsfysika, University of Leuven, Celestijnenlaan 200 D, B-3001 Leuven, Belgium}
\address[stuk]{STUK - Radiation and Nuclear Safety Authority, P.O. Box 14, Helsinki FIN-00881, Finland}
\address[petersburg]{Petersburg Nuclear Physics Institute, Gatchina, St-Petersburg, 188350, Russia}
\address[manc]{Nuclear Physics Group, Schuster Laboratory, Brunswick Street, University of Manchester, M13 9PL, UK}
\address[mainz]{Institut für Physik, Johannas Gutenberg-Universität, 55099 Mainz, Germany}
\begin{abstract}
A laser ion source is under development at the IGISOL facility, Jyväskylä, in order to address deficiencies in the ion guide technique. The %%@
key elements of interest are those of a refractory nature, whose isotopes and isomers are widely studied using both laser spectroscopic and %%@
high precision mass measurement techniques. Yttrium has been the first element of choice for the new laser ion source. In this work we %%@
present a new coupled dye-Ti:Sapphire laser scheme and give a detailed discussion of the results obtained from laser ionization of yttrium %%@
atoms produced in an ion guide via joule heating of a filament. The importance of not only gas purity, but indeed the baseline vacuum %%@
pressure in the environment outside the ion guide is discussed in light of the fast gas phase chemistry seen in the yttrium system. A %%@
single laser shot model is introduced and is compared to the experimental data in order to extract the level of impurities within the gas %%@
cell.
\end{abstract}
\end{frontmatter}
\section{Introduction}  % 1
In the early 1980's the ion guide technique was developed in Jyväskylä in order to overcome limitations related to the standard ISOL %%@
technique, namely the inability to produce refractory elements and a need for complicated ion source-target combinations. A full %%@
description of the IGISOL technique may be found in \cite{1aysto,2huikari} and references therein, however it will be briefly described %%@
here. A projectile beam impinges on a thin target and the reaction product nuclei recoil out into a chamber filled with a buffer gas, %%@
usually helium. The highly-charged ions slow down, thermalize in the gas, and continuously change their charge state until a significant %%@
fraction, 1-10\% reach a $1^+$ charge state. This fraction is transported out of the ion guide with the gas flow and is guided through a %%@
radio frequency sextupole ion guide (SPIG) before being injected into the mass separator. The key advantages of the IGISOL technique are %%@
the short release times of radioactive nuclei ($\approx$ ms timescales) and a chemical non-selectivity.
\par
	However, for certain reactions, there are two inherent deficiencies in this ISOL method, namely the lack of $Z$-selectivity and a poor %%@
efficiency. These drawbacks become more important in fission reactions involving the light-ion bombardment of heavy actinide targets. The %%@
plasma generated from the passing of the fission fragments through the stopping gas can lead to severe recombination losses. This %%@
deteriorates the ion guide efficiency.
\par
	In order to overcome the deficiencies arising from the plasma effect and to provide $Z$-selectivity an alternative approach to the %%@
IGISOL technique has been pioneered by the LISOL group at the University of Leuven \cite{3duppen,4yuri}. By allowing the recoiling product %%@
nuclei to neutralize a selective re-ionization process is achieved using lasers. This work has demonstrated the feasibility of combining %%@
high-power low-duty cycle lasers with the ion storage capability of a high pressure gas cell. By employing a pulsed primary beam with a %%@
time structure that is optimized for the evacuation time of the ion guide the initial plasma may be extracted out of the cell. The %%@
presence, therefore, of only weakly ionized plasma means that the gas flow rates to transport ions out of the ion guide can be %%@
significantly lower than in the standard IGISOL system. This permits heavier stopping gases to be used such as Ar, leading to a better %%@
stopping efficiency for recoiling nuclei. In comparison to the standard IGISOL however, the laser ion guides  suffer from longer delay %%@
times. It was demonstrated in these studies that the gas impurity plays a significant role down to the level of 1 ppb. Highly efficient %%@
ionization has been reached for Ni \cite{5yuri,6vermeeren} and several other refractory elements, such as Co, Rh and Ru \cite{7vanduppen}.
	\par
	At the IGISOL facility, a similar project is underway to combine the selectivity and efficiency of a laser ion source with the fast %%@
(sub-millisecond) release, and chemical non-selectivity of the ion guide technique \cite{8moore}. The fast release, extreme purity and %%@
substantially improved intensities will push existing spectroscopic experiments further from stability than is presently possible. Several %%@
techniques are being developed. One is similar to that pioneered by the Leuven group as discussed above. A second technique, developed in %%@
order to provide the highest selectivity, will ionize the neutral atoms after extraction from the gas cell, within the SPIG \cite{9moore}. %%@
This latter technique will not be discussed. In this paper we concentrate on issues leading to the development of the ``standard'' ion %%@
guide laser ion source.
\par
The first element for study with the new laser ion source is yttrium. This is a very challenging element due to the chemical reactivity of %%@
yttrium ions to any impurities within the gas cell. It has been chosen as a first case based on a need to extend a rich programme at the %%@
IGISOL facility of ground state nuclear structure studies in the refractory region \cite{10hager,11campbell,12kankainen}. From a physics %%@
point of view there is a threefold motivation to study yttrium: nuclear astrophysics interests, weak interaction physics (the superallowed %%@
beta decay of the $N=Z$ nucleus, $^{78}$Y) and nuclear structure physics. 
\par
In order to probe the properties of nuclei on either side of the valley of beta stability different reaction mechanisms are used to %%@
populate the isotopes of interest. Heavy-ion reactions can be used to produce nuclei close to the $N=Z$ line in regions that play a special %%@
role in nuclear astrophysics since the rapid-proton (rp) capture process passes right through them. The properties of these %%@
neutron-deficient nuclei, in particular the masses and beta decay half-lives, are needed to perform rp-process nuclear reaction network %%@
calculations. Due to the relatively low production yields using heavy-ion reactions at IGISOL the laser ion source aims to improve the %%@
yield of the nuclei of interest. Light-ion induced fission reactions are used in order to probe exotic nuclei on the neutron-rich side of %%@
the valley of stability. The resultant problem in this reaction is not the efficiency of producing the nuclei, rather the often %%@
overwhelming background from more abundantly produced isobars. In this case the laser ion source aims to provide a means of being more %%@
selective.
\section{Experimental set-up} % 2
In order to selectively ionize yttrium, resonance ionization schemes were developed involving two experimental components, the laser system %%@
of choice and an atomic beam source. Resonance ionization spectroscopy was performed off-line using a compact atomic beam unit, discussed %%@
in section \ref{sec:abu}. For a realistic study of laser ionization of yttrium a laser ion guide was loaned from the LISOL group at %%@
Louvain-la-Neuve. The details of this ion guide will be discussed in section \ref{sec:lisol}. In order to control the arrival time of the %%@
lasers such that evacuation and molecular formation within the ion guide may be studied, a mechanical shutter mechanism developed for this %%@
purpose is described in section \ref{sec:shutter}.

\subsection{ Laser system} % 2.1

\begin{figure}
\centering
\includegraphics[width=.95\linewidth]{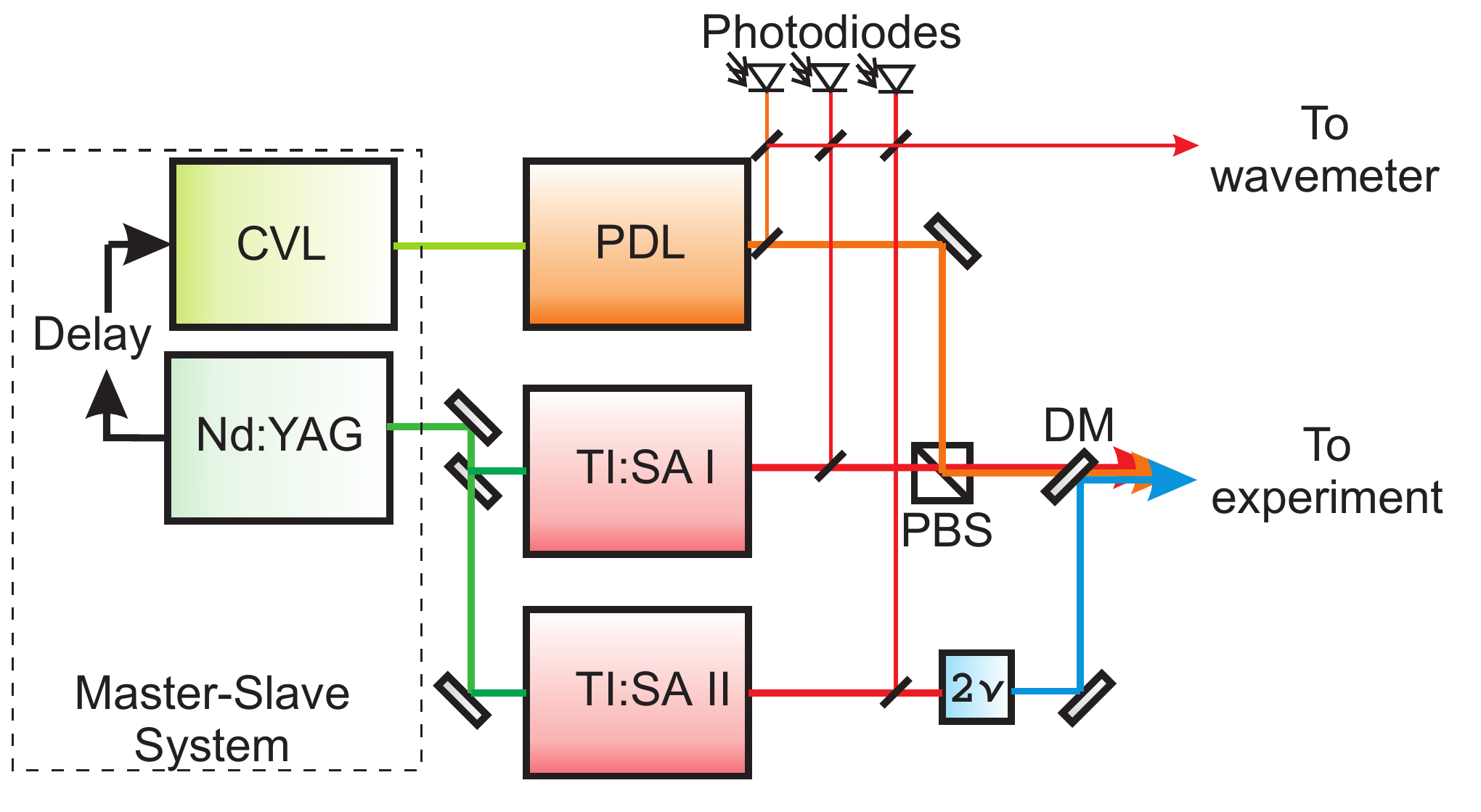}
\caption{ Laser system used for the resonant ionization of yttrium. PBS = Polarising Beamsplitter Cube, DM = dichroic mirror.}
\label{fig:lasersystem}
\end{figure}

The laser ion source facility at Jyväskylä employs a novel \textsl{twin} laser system, based on a combination of high-resolution titanium %%@
sapphire lasers and dye lasers, each with independent pump lasers. A more detailed description of the laser facility may be found in %%@
\cite{8moore}, however will be briefly described with emphasis on the lasers used specifically for the laser ionization of yttrium. Fig. %%@
\ref{fig:lasersystem} shows the optical layout of the laser system used in this work.
% Fig. 1. 
In the configuration shown in Fig. \ref{fig:lasersystem}, the primary (master) laser system, an all-solid-state system, consists of a high %%@
repetition rate (12 kHz) diode-pumped Nd:YAG laser (Lee Laser, LDP-200MQG) with up to 100 W average laser power in the second harmonic, %%@
pumping two tunable titanium sapphire (Ti:Sa) lasers. The average output power of Ti:Sa 1 was 1.5 W, while that of Ti:Sa 2 was 3 W. This %%@
increase in laser power for the second Ti:Sa is attributed to a double-sided pumping technique, described in detail in \cite{13kessler}. %%@
Each laser contains a Lyot filter and etalon for generation of narrow bandwidth laser radiation ($\approx$ 5 GHz). The temporal %%@
synchronization of the 30-50 ns pulses from the two lasers was achieved using remote-controlled Q-switches installed into each individual %%@
resonator. In order to access a variety of first excited states in yttrium, second harmonic generation with a beta-barium borate (BBO) %%@
crystal extended the fundamental IR range of the Ti:Sapphire laser into the blue region of the electromagnetic spectrum. With approximately %%@
3 W fundamental power from Ti:Sa 2 a typical frequency doubled power of $\approx$ 400 mW was achieved.
\par
The second laser system (slave) consists of a high repetition rate copper vapour pump laser (Oxford Lasers LM100X(KE)) with up to 45 W %%@
average laser power. Two fundamental lasing wavelengths are attainable with this laser, 511 nm (green) and 578 nm (yellow), with an %%@
approximate power ratio of 2:1. The 511 nm laser light is used to pump a pulsed dye laser (Spectra Physics PDL-3) running with a DCM dye %%@
dissolved in methanol, enabling an output wavelength range of 610-660 nm. The typical output power during this work was 400 mW at around %%@
640 nm. This is lower than expected based on the typical powers achieved at ISOLDE using a copper vapour laser pumped dye laser, however %%@
the speed of the JYFL dye circulators was optimized for low repetition rate pumping only \cite{14tordoff}. As of the time of writing, the %%@
circulators have been upgraded and a typical pulsed dye laser power is now close to 1 W at 640 nm. The linewidth and temporal pulse length %%@
of the dye laser are 1 GHz and 20 ns, respectively.
\par
\begin{figure}
\centering
\includegraphics[width=.95\linewidth]{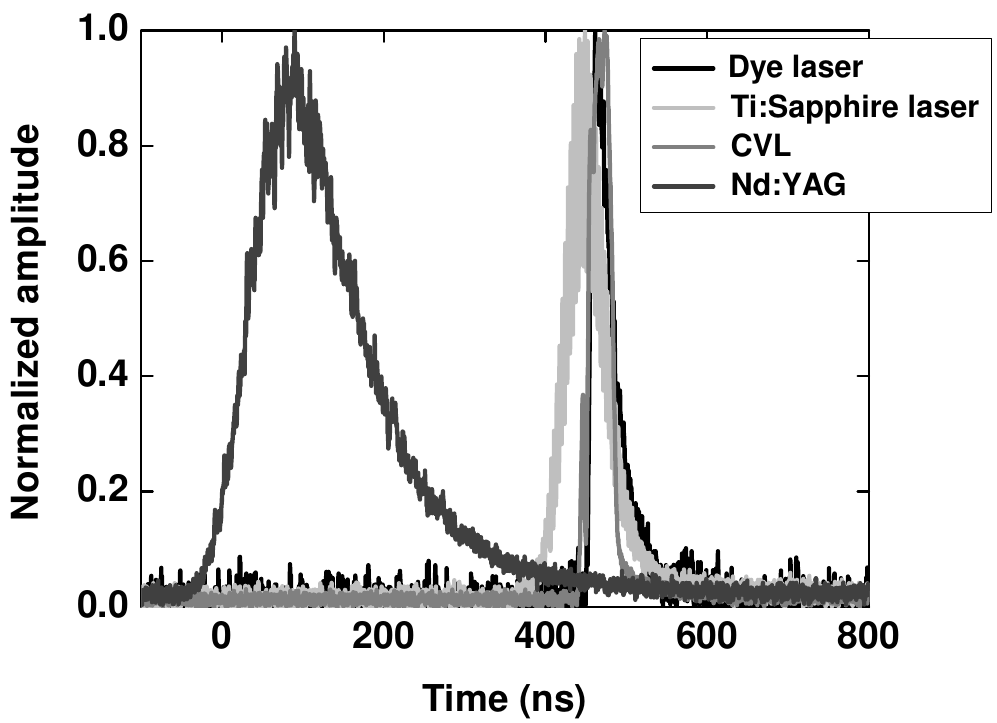}
\caption{ Temporal profiles of the two pump lasers, a single Ti:Sapphire laser and a pulsed dye laser.}
\label{fig:templaser}
\end{figure}

Fast silicon photodiodes (Roithner LaserTechnik SSO-PD-Q-0.25-5-SMD) are used to monitor the temporal overlap of the individual lasers. %%@
Fig. \ref{fig:templaser} shows the temporal profiles of the Nd:YAG, the copper vapour laser, a single Ti:Sapphire laser and the dye laser. 
% Fig. 2. 
In the case of a free-running system the lasers do not fire synchronously due to the different build-up times needed to attain a population %%@
inversion within the two types of gain media. This would result in the inability to achieve a stepwise resonant excitation (lifetimes of %%@
typical atomic excited states are a few ns). While the Q-switches are used to attain a temporal overlap between the Ti:Sapphire lasers, the %%@
copper vapour laser can be triggered by the master clock of the Nd:YAG laser via a delay such that the dye laser pulse is synchronized with %%@
the Ti:Sa pulses.
\par
Approximately 1\% of the output power from each laser is used for monitoring the temporal overlap and wavelength. An optical fibre couples %%@
the individual lasers into a commercial wavemeter (High Finesse, WS-6) for wavelength determination. The main beams from the dye laser and %%@
Ti:Sa 1 are coupled together spatially using a polarizing beamsplitter cube (PBS, see Fig. \ref{fig:lasersystem}), and are then overlapped %%@
with the frequency doubled laser light of Ti:Sa 2 using a dichroic mirror (DM). The laser beams  are sent overlapped both temporally and %%@
spatially either to an atomic beam unit, described in the following section, or to the IGISOL as described in section \ref{sec:experiment} %%@
for experiments.
\subsection{ Atomic beam unit} %2.2.
\begin{figure}
\centering
\includegraphics[width=.95\linewidth]{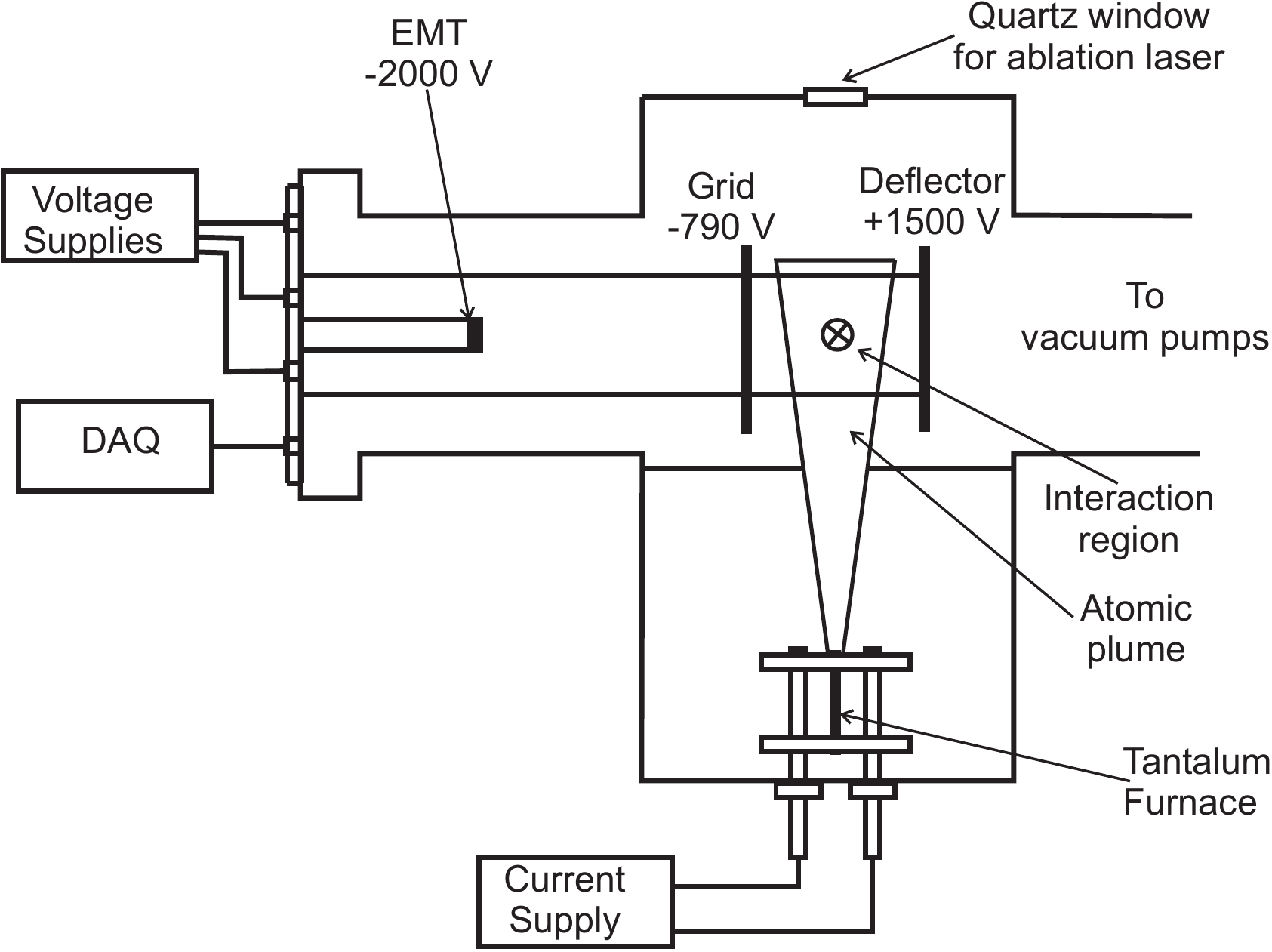}
\caption{Schematic diagram of the atomic beam unit used for testing resonant ionization schemes. The principle is described in the text. %%@
The applied voltages have been optimized experimentally to provide the maximum ion counting rate on the EMT.}
\label{fig:abu}
\end{figure}

\label{sec:abu}
An atomic beam unit (ABU) was used to test the relative efficiency of several ionization schemes for atomic yttrium. A schematic diagram of %%@
the ABU is shown in Fig. \ref{fig:abu}. 
% Fig. 3. 
The operational principle is based on a crossed-beams technique. A sample of yttrium was loaded into a tubular tantalum furnace with a %%@
length of 50 mm, an outer diameter of 2 mm and a thickness of 0.5 mm. The sample is vaporized under vacuum ($\approx 1 \cdot 10^{-5}$ mbar) %%@
using a joule heating current of approximately 50 A. An atomic plume of yttrium passes through the laser interaction region and is stepwise %%@
resonantly excited and ionized. An electrostatic field, generated by a deflector-grid system using the potentials as shown in Fig. %%@
\ref{fig:abu}, is used to extract the ions towards an electron multiplier tube (EMT) (ETP Electron Multipliers, Model AF150H). At an %%@
operating potential of -2 kV the multiplier gain is specified to approximately $3 \cdot 10^7$. Fast negative ion signals from the EMT are %%@
amplified, discriminated from the background noise and shaped for counting using a PC counter card. A labview-based computer control %%@
program is used to scan the laser wavelengths while simultaneously recording the ion signal. In order to reduce background counts from the %%@
fast edges of the Q-switches and the thyratron of the CVL laser, the signal is gated on the arrival time of the ions after the lasers fire.
\par
In order to study more refractory elements that cannot be extracted from the oven via joule heating, the experimental setup can be modified %%@
in order to produce a pulsed atomic plume of the species of interest. In this case the tantalum oven is removed in favour of a metallic %%@
``boat'' mounted on an electric motor with a variable rotation speed. A quartz window on the top of the ABU enables the delivery of a low %%@
repetition rate (20 Hz) 1064 nm Nd:YAG laser operating with a typical pulse energy of 25 mJ, to be used for laser ablation %%@
\cite{15tordoff}. In this running mode the ion signal is additionally gated on the trigger pulse of the ablation laser.
\subsection{Laser ion guide} % 2.3
\begin{figure}
\centering
\includegraphics[width=.95\linewidth]{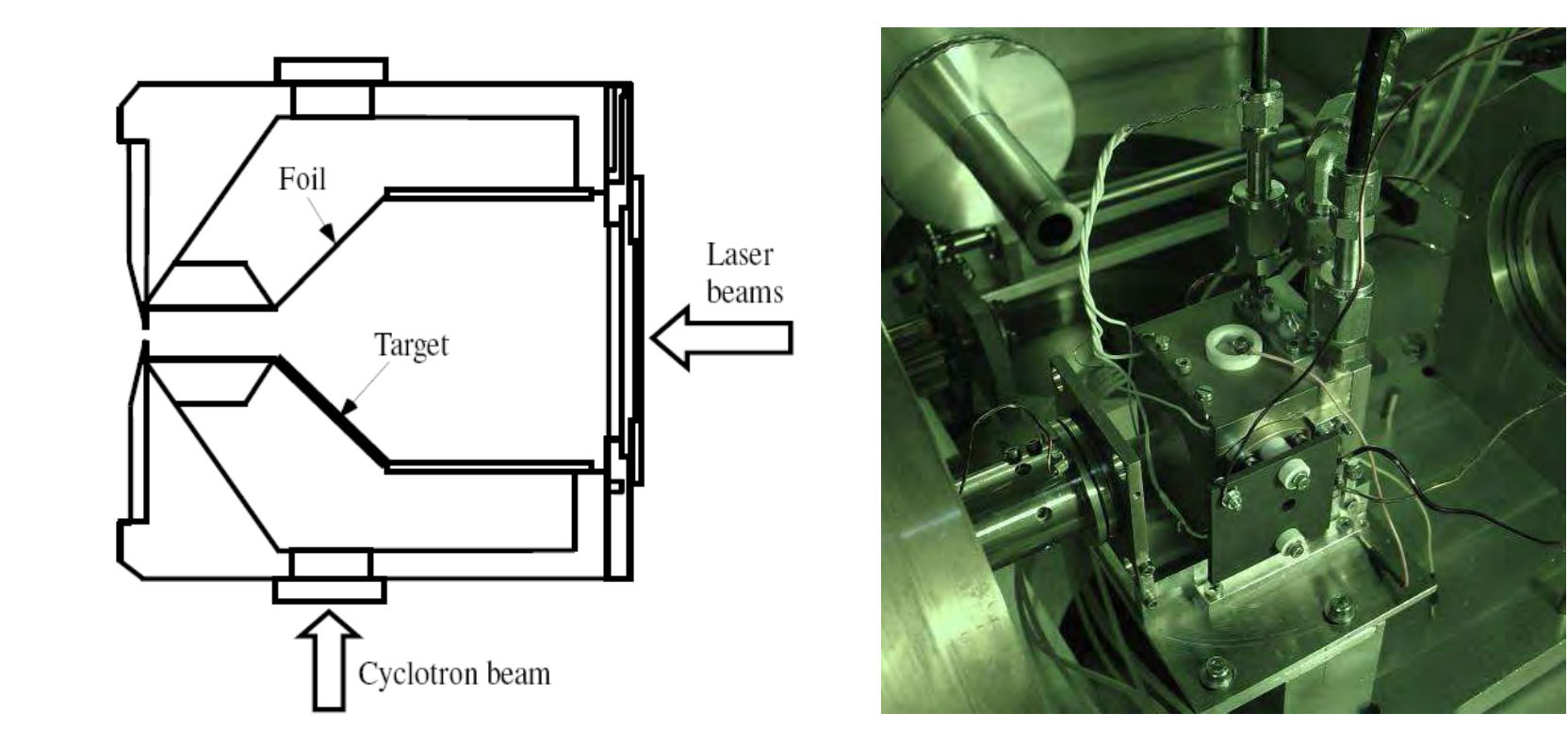}
\caption{Schematic view of the laser ion source loaned by the LISOL group at Louvain-la-Neuve (left) and a picture of it installed in the %%@
IGISOL chamber (right). In the picture the lasers enter from the right, and the SPIG is seen directly in front of the ion guide exit hole. %%@
The beam pipe towards the beam dump can be seen in the top of the picture.}
\label{fig:lisol}
\end{figure}

\label{sec:lisol}
In this work a laser ion guide loaned by the LISOL group at Louvain-la-Neuve was used and is shown in Fig. \ref{fig:lisol}.
% fig4
The body of the ion guide is made from brass, and all parts are sealed with indium enabling heating of the whole guide up to 120 °C. The %%@
volume of the gas cell is split into two parts, the main volume into which a primary beam enters and where a filament can be mounted and a %%@
channel of 10 mm in diameter and 26 mm in length, termed the ionization channel. The conductance of the exit hole of 0.5 mm diameter in %%@
helium gas is equal to 0.112 l/s. An evacuation time for the whole guide of 480 ms can be estimated by dividing the stopping chamber volume %%@
by the exit-hole conductance. The evacuation time of the ionization channel is similarly calculated and is 18 ms. Compared to the ion %%@
guides used in JYFL these extraction times are considerably longer due to the smaller exit hole, however an increase in time is necessary %%@
in order to achieve neutralization of the recoiling nuclei. Atoms of different elements can be produced inside the cell by the resistive %%@
heating of a corresponding filament, either made directly from the chosen element, or with material evaporated onto the surface of a %%@
filament. In this work, an yttrium filament of thickness 3.75 mg/cm$^2$ cut into a bowtie configuration with a diameter of 3 mm at the %%@
narrowest part was mounted within the guide towards the rear (not shown in Fig. \ref{fig:lisol}). Laser light enters the cell %%@
longitudinally through a sapphire window in the rear, ionizing neutral atoms along the axis. The diameter of the laser beam is brought to a %%@
gradual focus of approximately 3 mm near the exit hole. 
\subsection{Control of the laser to ion guide arrival time} %2.4
\begin{figure}
\centering
\includegraphics[width=.95\linewidth]{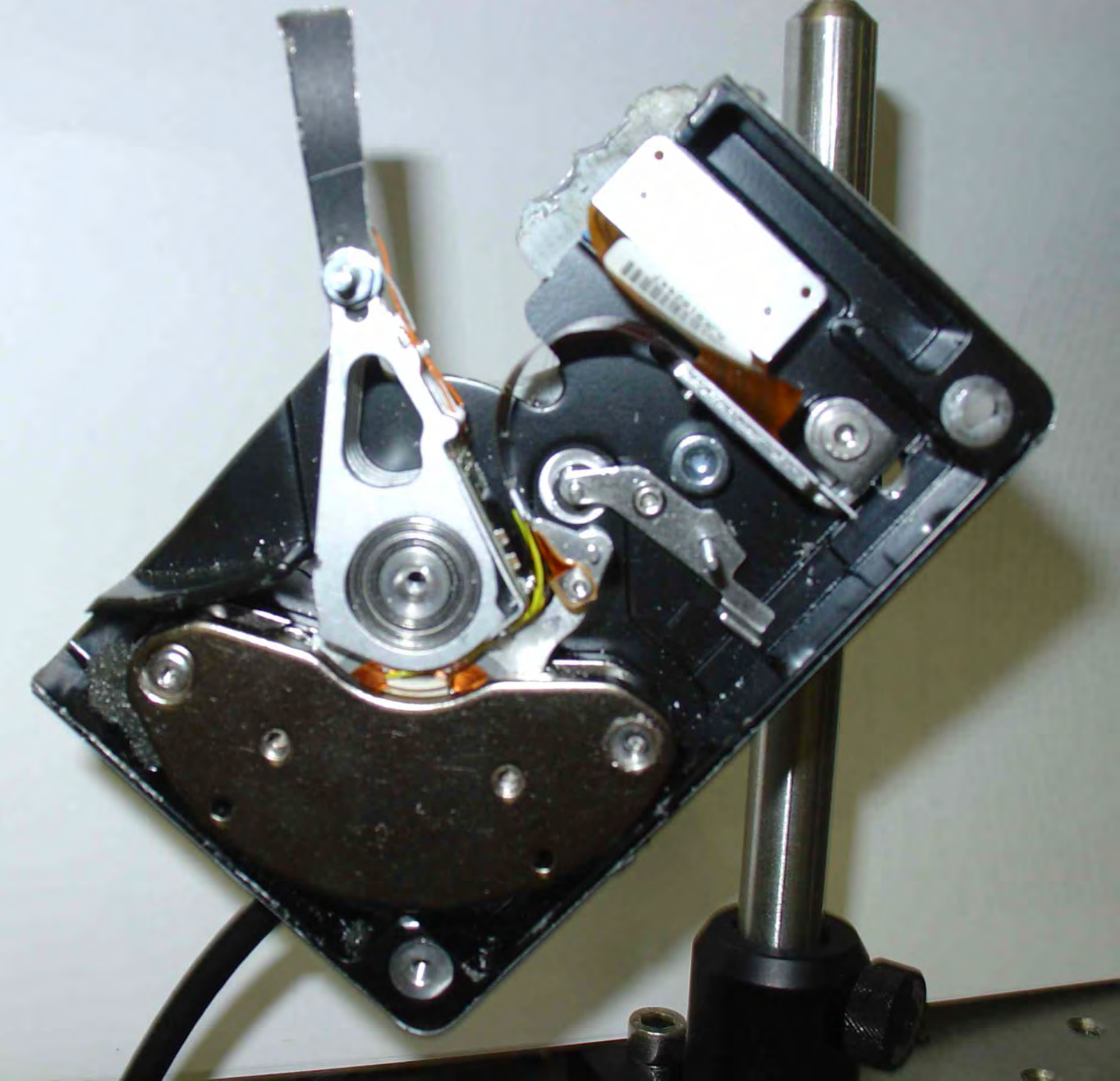}
\caption{ A photograph of the modified hard disk with extended readout arm used for mechanical control of the laser arrival time at the ion %%@
guide.}
\label{fig:shutter}
\end{figure}

\label{sec:shutter}
The temporal control of the spatially overlapped lasers is mandatory for the investigation of the evacuation time and the timescales of any %%@
subsequent molecular formation within the ion guide. For low repetition rate laser systems with a duty cycle lower than the evacuation time %%@
of the ion guide a DAQ trigger can be used. However, this is not possible for high repetition rate systems such as that used in this work %%@
and therefore a fast shutter mechanism was designed to provide the required timing information. As the required time resolution is %%@
restricted by the bin width of the data readout system ($\approx$ 0.5 ms) a mechanical solution is still feasible and preferred over a fast %%@
(ns) but rather expensive light switch in a typical polarizer/analyzer setup in combination with a pockels cell as proposed in %%@
\cite{16mccann}. In this experiment a modified version of the mechanical shutter design detailed in \cite{17singer} was used. A %%@
conventional broken hard-drive disk was dismantled and the readout arm of the disk was connected to an electrical switching circuit %%@
providing a shutter current of roughly 500 mA. The shutter arm was extended with a thin aluminium plate to decrease the shutter time (Fig. %%@
\ref{fig:shutter}). 
\par
\begin{figure}
\centering
\begin{minipage}[t]{0.49 \linewidth}
\includegraphics[width=.95\linewidth]{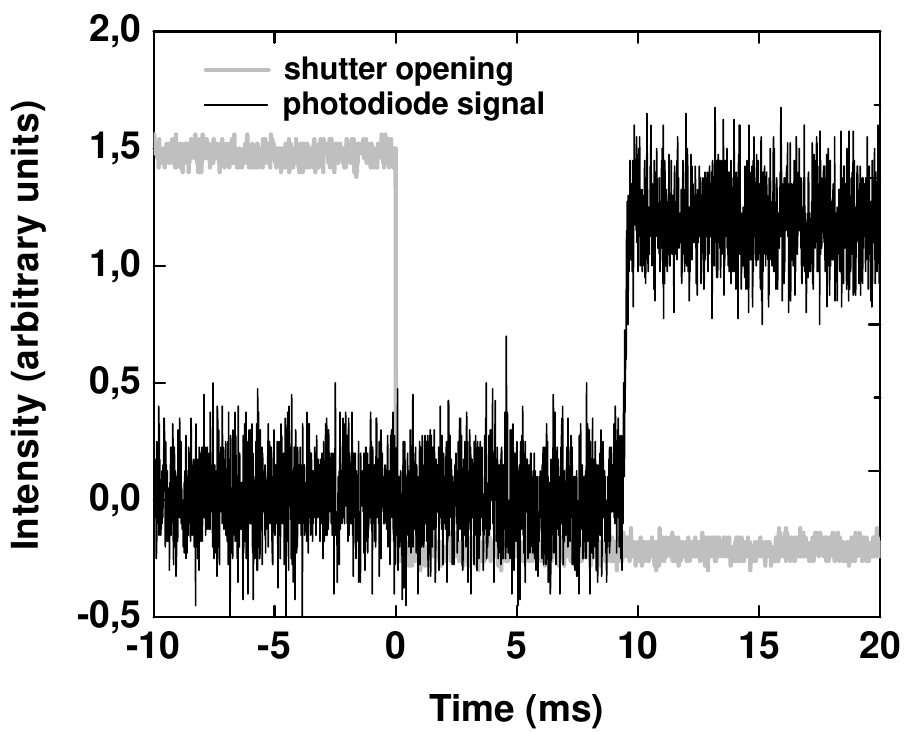}
\end{minipage} \hfill
\begin{minipage}[t]{0.49 \linewidth} 
\includegraphics[width=0.95 \linewidth]{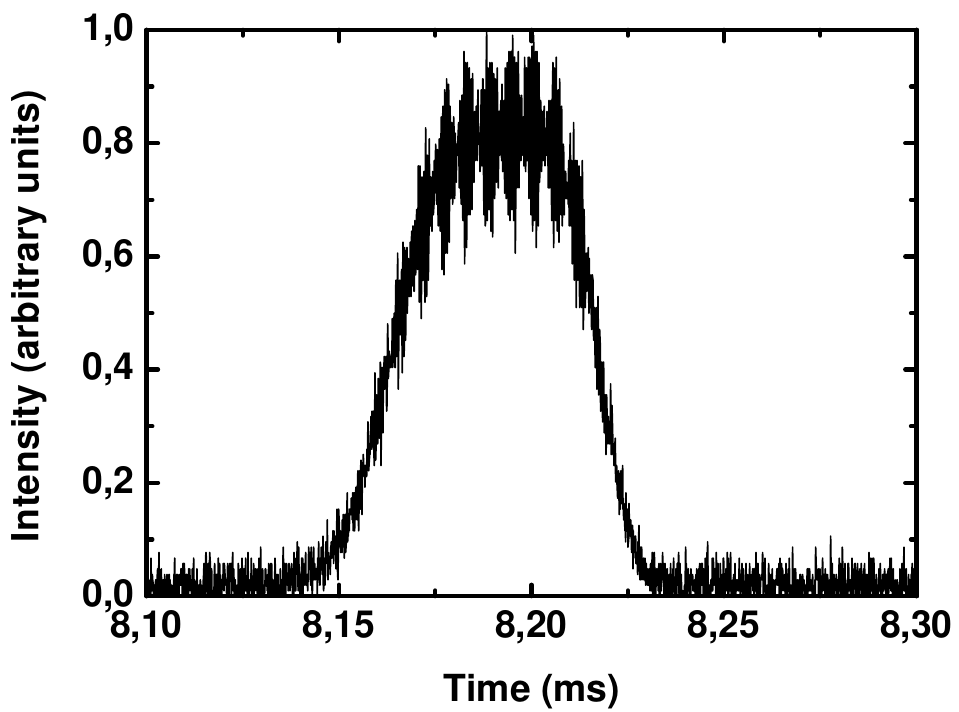} 
\end{minipage}
\caption{ The delay between the TTL for the single-arm shutter and the detection of the helium-neon laser (left). With the double shutter %%@
system in place the control of the temporal overlap is of the order of 100 microseconds (right).}
\label{fig:shutterspecs}
\end{figure}

The direction of the feeding current could be switched by using standard TTL logic. The performance of the shutter was tested using a %%@
continuous wave helium-neon laser beam focused in the same manner as that of the Ti:Sapphire laser beams. The light was detected by a fast %%@
photodiode behind the shutter. A typical delay time of 5-14 ms between the TTL signal and the shutter mechanism was observed depending on %%@
the alignment of the shutter with the laser beam, with an overall jitter of 50 microseconds. A shutter time of roughly 20 microseconds was %%@
achieved. The long delay of the shutter does not allow for laser ``on'' timescales of less than 15 ms. Therefore, to reach the single laser %%@
shot level for a high repetition rate laser system (each laser pulse is separated by 100 microseconds in a 10 kHz system) a double shutter %%@
system was recently designed. In this approach two shutters were put in series into the laser beam so that shutter 1 opens the laser path %%@
and shutter 2 blocks the path. The damping of the movement of the shutters was improved to reduce the jitter of the shutter mechanism. By %%@
choosing an appropriate TTL logic a pulse of laser radiation of 100 microseconds was achieved as illustrated in Fig. %%@
\ref{fig:shutterspecs}.
% Fig. 5
% fig6
\section{Results and discussion} %3
\label{sec:experiment}
The off-line development described in this paper has led to a number of results which will be discussed within the following individual %%@
subsections. Section \ref{sec:ris} details the work carried out within the atomic beam unit on yttrium using the mixed dye-Ti:Sapphire %%@
laser system to provide a suitable resonance ionization scheme to be tested in the laser ion guide. The initial work done in the LISOL gas %%@
cell is detailed in section \ref{sec:06}, including first studies on the gas phase chemistry performed in March 2006. Section \ref{sec:07} %%@
highlights the results obtained one year later after similar gas purification steps used in the previous experiment, and furthermore %%@
discusses the effect of introducing impurities into the IGISOL chamber via a controlled leak valve.
\subsection{Development of a laser ionization scheme for yttrium} %3.1
\begin{figure}
\centering
\includegraphics[width=.95\linewidth]{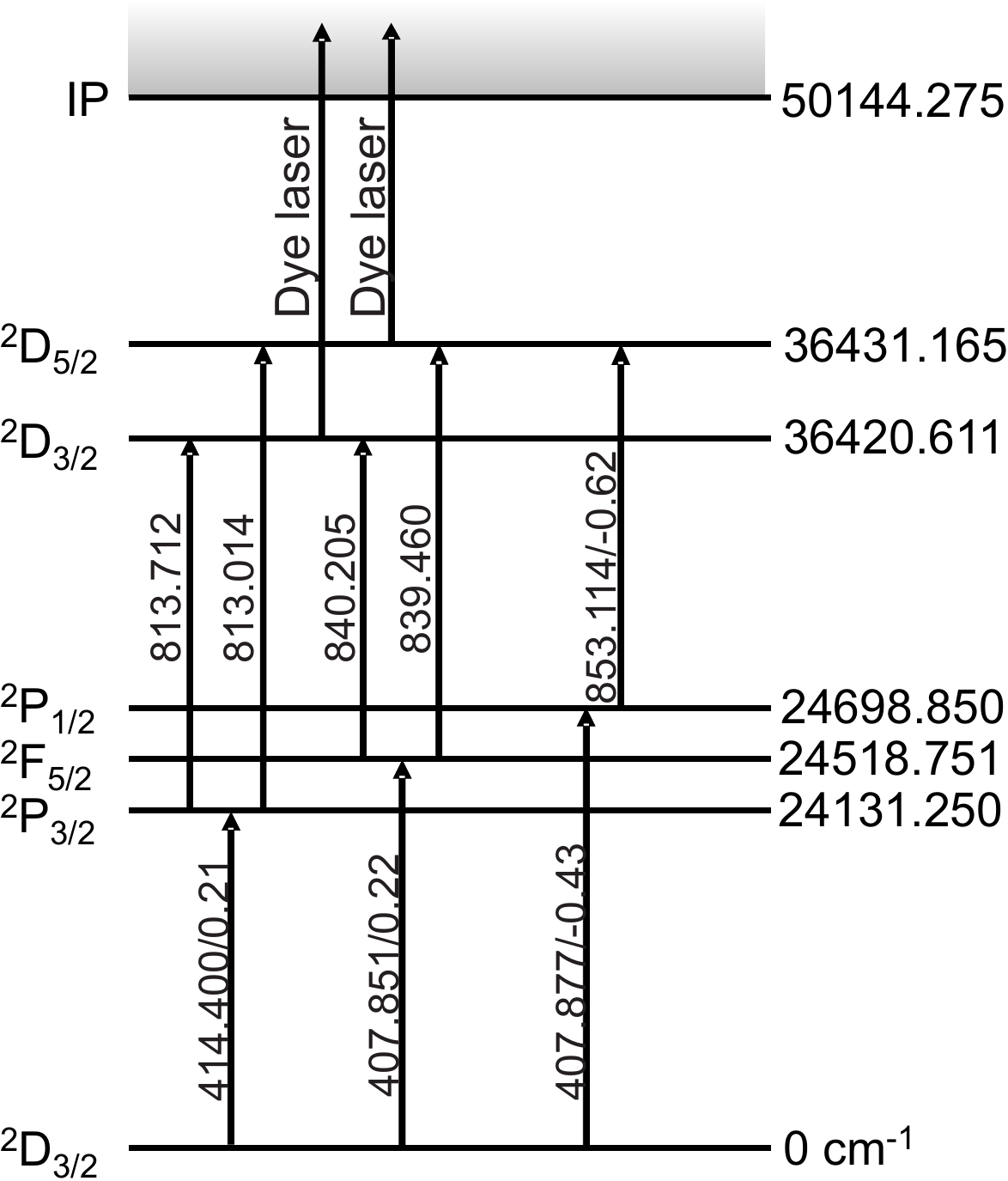}
\caption{ Relevant atomic energy levels in yttrium and the transitions investigated in this work.}
\label{fig:risschemes}
\end{figure}

\label{sec:ris}
The search for an appropriate resonant ionization scheme is of crucial importance prior to working with a laser ion source. Ideally, as in %%@
the case of the work discussed in this article, off-line searches for new schemes can be done without the need for an ISOL facility. The %%@
most important priority is the comparison of the efficiencies of the ionization schemes. For details on the theory behind resonance %%@
ionization and related experiments involving this technique the reader is referred to references \cite{18letokhov,19hurst}. An atomic %%@
energy level scheme showing the transitions investigated in this work is shown in Fig.  \ref{fig:risschemes}.
% Fig. 7 
\par
The atomic excitation energy of the levels labeled on the right of Fig. \ref{fig:risschemes} is given in units of cm$^{-1}$. The energy %%@
levels are identified on the left side and are given in the Russel-Saunders-Coupling notation with the atomic structure labeled as %%@
$^{2S+1}L_J$, where $S$ is the total spin, $L$ the total orbital angular momentum and $J$ the quantum number of the spin-orbit interaction. %%@
The wavelengths of the transitions between the levels are denoted in units of nm and are given as wavelength in vacuum. Next to the %%@
wavelength, the log-gf (transition strength) is listed if known from literature \cite{20kurucz,21hirata}.
\par
The first resonant transition step was realized using a frequency-doubled Ti:Sapphire laser, while the second resonant step was driven by %%@
the fundamental output of a second Ti:Sapphire laser. A non-resonant ionization step was achieved by exciting the atoms across the %%@
ionization potential (IP), 50144.275 cm$^{-1}$, with the frequency doubled light from the first transition. The dye laser was used to scan %%@
across the IP to search for the existence of possible auto-ionizing levels. These levels can then be used in a third resonant ionizing step %%@
and provide an enhancement in the cross section probability for ionization compared to that achieved for a simple non-resonant step. 
\par
A comparison between four different Ti:Sapphire ionization schemes tested in the atomic beam unit are shown in table \ref{tab:ris}.
% Table 1
\begin{table}
\centering
\caption{ Comparison of RIS schemes performed in the atomic beam unit.}
\begin{tabular}{ccccc}
\hline
$\lambda_1$(nm) &	$\lambda_2$(nm) &	$P_2$ (W)	 & Count rate (s$^{-1}$)	& Count rate / $P_2$ \\ \hline
407.85 &    839.46 &	1.9 &	4000 &   2105 \\ \hline
407.85 &	840.21 &	1.4 &	2700 &	1929 \\ \hline
414.40 &	813.71 &	1.1 &	4000 &	3636 \\ \hline
414.40 &	813.01 &	1.4 &	4200 &	3000\\ \hline
\end{tabular}
\label{tab:ris}
\end{table}
A constant current of $\approx$ 53 A was supplied to the oven for the duration of the tests. In all cases the fundamental Ti:Sapphire laser %%@
power prior to frequency doubling for the first step was 2.1 W, leading to 300 mW in the second harmonic. This was high enough to achieve %%@
saturation for all first step transitions. The maximum power of the second steps varied between 1.1 W and 1.9 W. The count rate detected on %%@
the electron multiplier tube is shown in column 4 of table 1. It is unlikely that the second step transition is being saturated and %%@
therefore it is more informative to show the ratio of the observed count rate to the power of the second step. This is calculated in the %%@
final column. Even though the laser power in the second step was lower for the $\lambda_2$ = 813.71 nm transition compared to the 813.01 nm %%@
transition, a higher ratio of count rate to second step power was achieved. This infers a larger transition strength of the 813.71 nm %%@
transition. A fifth scheme, $\lambda_1$ = 404.877 nm and $\lambda_2$ = 853.114 nm, was tested in the ion guide however showed no %%@
improvement and therefore will not be discussed further.
\par
\begin{figure}
\centering
\begin{minipage}[t]{0.49 \linewidth}
\includegraphics[width=.95\linewidth]{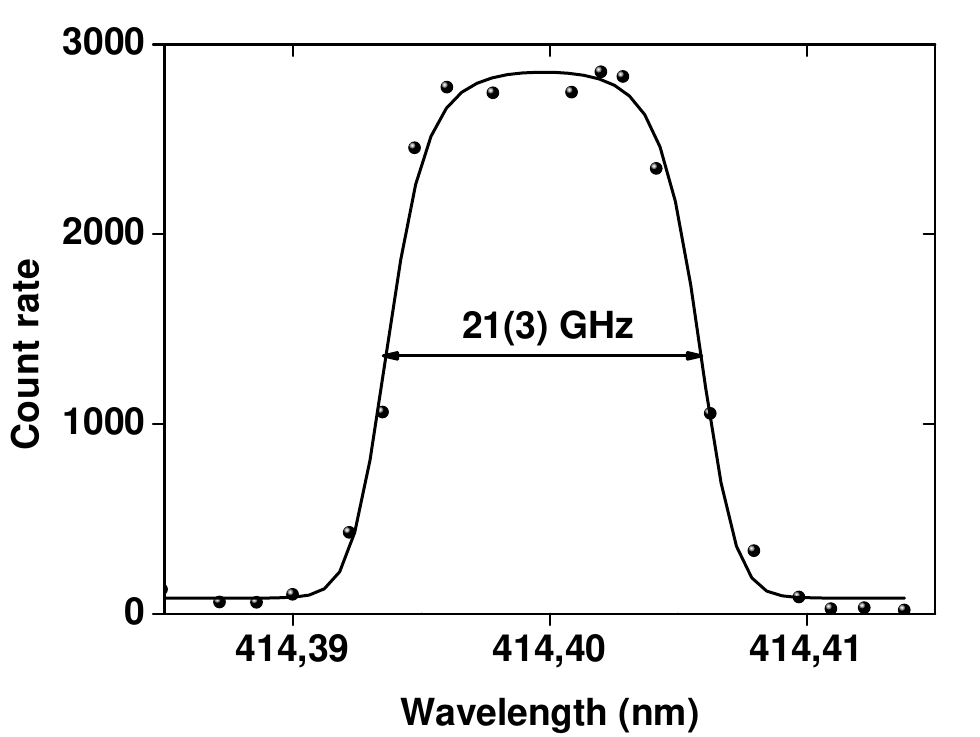}
\end{minipage} \hfill
\begin{minipage}[t]{0.49 \linewidth} 
\includegraphics[width=0.95 \linewidth]{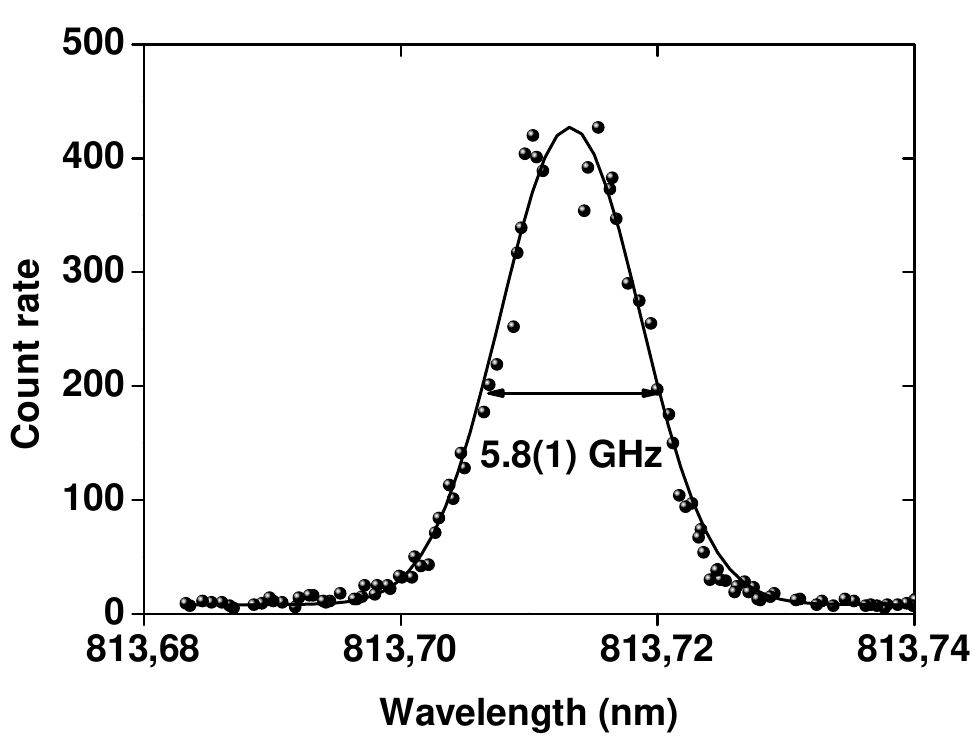} 
\end{minipage}
\caption{ Resonant laser wavelength scans of the most efficient first and second step transitions. The full-width at half-maximum (FWHM) %%@
value is shown.
}
\label{fig:laserscan12}
\end{figure}

	A graph showing the laser wavelength scans of the two resonant transitions from the optimum scheme of table 1 is shown in Fig. %%@
\ref{fig:laserscan12}.
% fig8
A fit of the form
\begin{align} 
y(\nu)=y_0+A_0 \cdot \frac{S(\nu)}{2(S(\nu)+1)}
\label{eq:1}
\end{align}
with
\begin{align} 
S(\nu)=S_0 \cdot e^{-0.5((\nu-\nu_0)/w)^2}
\label{eq:2}
\end{align}
was applied for the first step transition acknowledging a saturated lineshape profile dominated by the Gaussian  spectral profile of the %%@
laser radiation. $S(\nu)$ denotes the frequency dependent saturation parameter, $y_0$ a signal offset and $A_0$ the signal amplitude. The %%@
factor $S_0$ describes the resonant saturation magnitude for a two-level system. This provides a good approximation due to the high level %%@
of saturation of the first transition compared to the second and final transitions. Fitting of the data yields a value of 49(15) for $S_0$, %%@
supporting the conclusion that the first step is indeed saturated. The second step was fitted with a non-saturated Gaussian function %%@
corresponding to $S_0 \ll 1$ in Eq. \eqref{eq:2}.
\par
\begin{figure}
\centering
\begin{minipage}[t]{0.49 \linewidth}
\includegraphics[width=.95\linewidth]{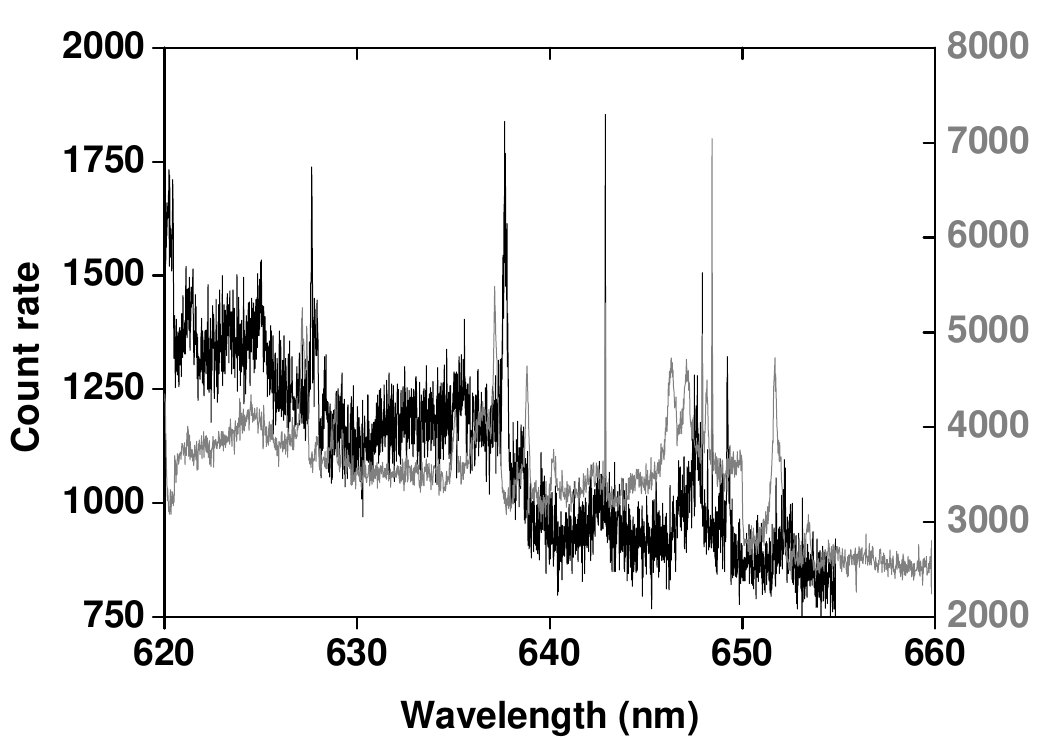}
\end{minipage} \hfill
\begin{minipage}[t]{0.49 \linewidth} 
\includegraphics[width=0.95 \linewidth]{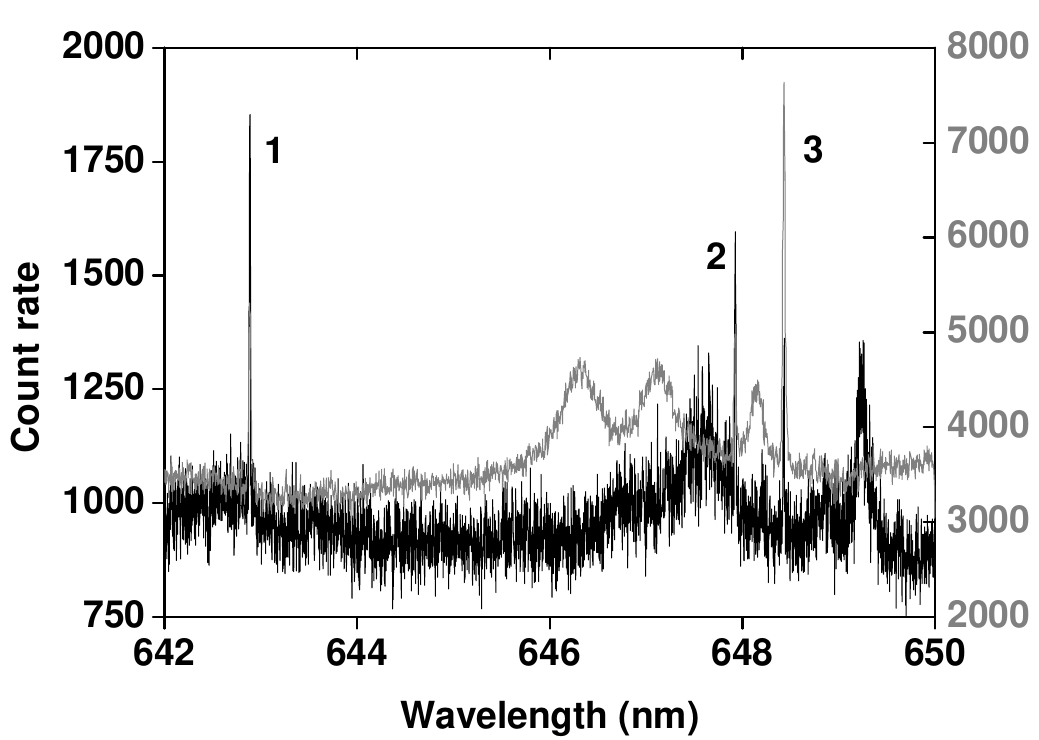} 
\end{minipage}
\caption{ The left hand figure shows the dye laser scan, while the spectrum on the right is a magnified part highlighting the three narrow %%@
resonances.
}
\label{fig:dyescan}
\end{figure}

	The pulsed dye laser (PDL) was used to search for auto-ionizing levels. Two wavelength scans were made over the range of the available %%@
dye (DCM, 610 - 660 nm) starting from the excitation levels populated by the two second step Ti:Sapphire transitions at 36420.611 cm$^{-1}$ %%@
and 36431.165 cm$^{-1}$. A summary of the lasers used for the different transitions and the laser powers measured at the entrance to the %%@
atomic beam unit is shown in table \ref{tab:ris2}.
%Table 2
\begin{table}
\centering
\caption{ The mixed Ti:Sapphire - dye transitions and laser powers obtained at the entrance to the atomic beam unit.}
\begin{tabular}{cccc}
\hline
Step    &	  $\lambda$ (nm)	  &   Power at abu (mW)       &    	Laser \\ \hline
1       &	414.40             	& 240	                     & Ti:Sa 2$^\text{nd}$ harmonic \\ \hline
2       &	813.71              & 	750	                   & Ti:Sa \\ \hline
2       &	813.01	            & 1050                     &	Ti:Sa \\ \hline
3       &	619-655             &	200-400                  &	Dye (DCM) \\ \hline
\end{tabular}

\label{tab:ris2}
\end{table}
Fig. \ref{fig:dyescan} shows the resultant dye laser spectrum starting from excitation levels of 36431.165 cm$^{-1}$ (black line, left %%@
scale) and 36420.611 cm$^{-1}$ (grey line, right scale), respectively. 
% fig9
A spectrum of resonances of differing intensities and widths was observed. The numerous broad structures were identified as auto-ionizing %%@
states. This was confirmed by blocking the second excitation step. If the signal drops to the background level then the second step is %%@
needed to perform the triple resonant ionization process. Three narrow resonances were also identified in both third step dye laser scans %%@
overlapping at exactly the same wavelength. These are labeled as 1, 2 and 3 in the more detailed view of the dye laser spectrum shown in %%@
the right hand figure of Fig. \ref{fig:dyescan}. It was concluded after further investigation that these resonances correspond to the dye %%@
laser driving a second step transition from 24131.250 cm$^{-1}$ to levels at 39686.0 cm$^{-1}$, 39565.1 cm$^{-1}$ and 39553.0 cm$^{-1}$ %%@
(resonances 1, 2 and 3 respectively), not shown in Fig. \ref{fig:risschemes}. The atomic levels at 39686.0 cm$^{-1}$ and 39565.1 cm$^{-1}$ %%@
are previously known, while the level at 39553.0 cm$^{-1}$ is newly discovered.
\par
\begin{figure}
\centering
\begin{minipage}[t]{0.49 \linewidth}
\includegraphics[width=.95\linewidth]{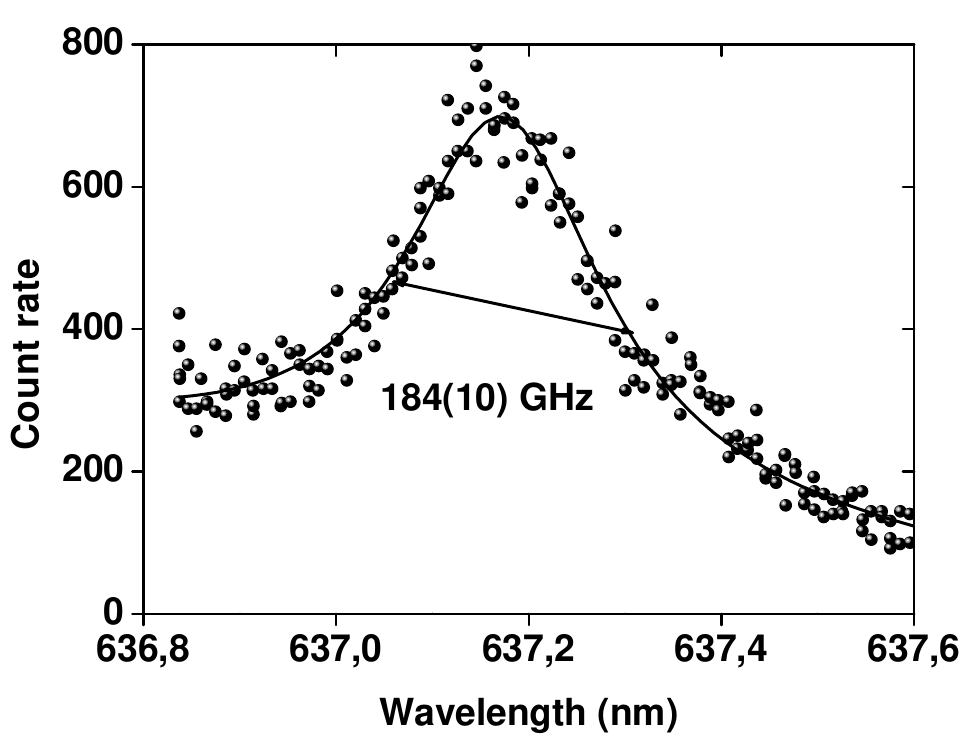}
\end{minipage} \hfill
\begin{minipage}[t]{0.49 \linewidth} 
\includegraphics[width=0.95 \linewidth]{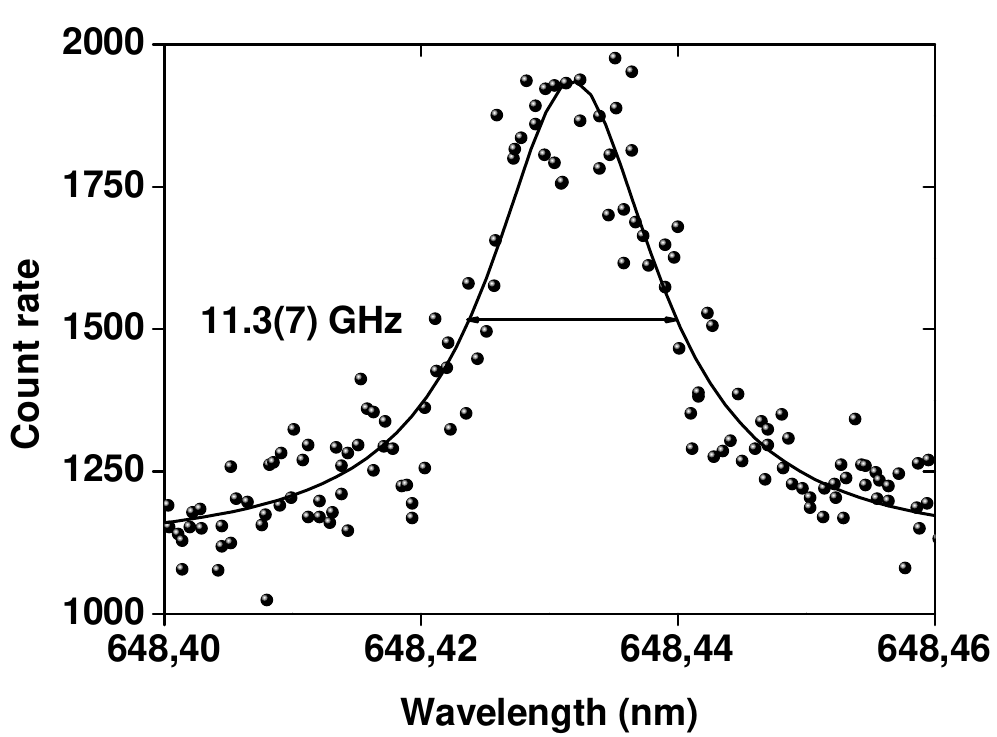} 
\end{minipage}
\caption{ The left hand figure shows a dye wavelength scan and the corresponding lineshape profile of a typical auto-ionizing state in %%@
yttrium, while the resonance on the right corresponds to a scan and lineshape fit of the bound state resonance peak 3 in Fig. %%@
\ref{fig:dyescan}.
}
\label{fig:dyescandetail}
\end{figure}

The difference in linewidth between the second step excitation driven by the dye laser and that of a standard auto-ionizing resonance is %%@
shown in Fig. \ref{fig:dyescandetail}. 
% fig10
Due to the non-saturated transitions both resonances were fitted by a Lorentzian lineshape profile on a linear background of the form
\begin{align} 
y(\nu)=y_0+b(\nu-\nu_c)+A \cdot \frac{w}{4(\nu-\nu_c)^2+w^2}
\label{eq:3}
\end{align}	
	The linear background has a slope $b$  and a constant offset $y_0$ at the centroid frequency $\nu_c$, while the last term is a %%@
Lorentzian with a full-width at half-maximum (FWHM)of $w$ and amplitude $A$. The auto-ionizing resonance has a FWHM  of 184(10) GHz. This %%@
can be compared to a FWHM of 11.3(7) GHz in the second scan, more typical of a bound state resonance. 
\par
Table \ref{tab:ris3} details the effects of blocking the different laser transitions, using the dye laser driving the new second step %%@
transition from the atomic level at 24131.250 cm$^{-1}$ to the level at 39553.0 cm$^{-1}$. 
\begin{table}
\centering
\caption{ The effect of blocking different lasers on the ion count rate.}
\begin{tabular}{cccccc}
\hline
&  Ti:Sa	
&	 Ti:Sa
&	Ti:Sa 
&	Dye 
&	All \\

&  	1$\text{st}$ step 
&	1$\text{st}$  step
&	2$\text{nd}$  step
& 2$\text{nd}$  step
&	 lasers \\

&  blocked & only & blocked & blocked & available \\ \hline
Count rate (ions/s) & 0	 &	10   &	600   &	1200	 &      1800 \\ \hline
\end{tabular}
\label{tab:ris3}
\end{table}
It is clear that the first step in the scheme is pivotal. When this transition is blocked the ion signal disappears. If only the blue %%@
transition is used it is possible to have a double-blue non-resonant transition after the first step, though this is very weak, only 0.5\% %%@
of the total count rate. By reducing the temperature of the oven this signal disappears and so these ions are indeed created via %%@
interaction with the laser beam. By including the second resonant Ti:Sapphire transition approximately 2/3 of the total ion signal is %%@
achieved. The main ionization pathway used in this case is then blue-IR-blue, where the final blue step is non-resonant. If the dye laser %%@
is used alone to drive a second step transition then approximately 1/3 of the total signal, 600 ions/s, is reached. In this scenario the %%@
possible ionization schemes are blue-red-blue and blue-red-red. When all three lasers are allowed to interact with the yttrium atoms, %%@
additional blue-IR-red and blue-red-IR pathways are possible, resulting in a total ion signal of 1800 ions/s.
\par
The reason for the different rates depending on which second step laser is blocked may be due to different transition strengths, or to any %%@
efficiency differences in pumping using a dye or Ti:Sapphire laser. Future investigations on the latter possibility are needed and an %%@
atomic system will have to be found that enables the same transition wavelength to be used. In this manner the intrinsic differences %%@
between the laser systems, such as the pulse length (10 ns for a dye laser, 50 ns for a Ti:Sapphire laser) may be studied as a function of %%@
the ionization efficiency. However, for the purpose of a laser ion source, the high peak ion count rate achieved with this mixed %%@
dye-Ti:Sapphire scheme has led to the use of the twin laser system for the majority of the following studies in this article. Indeed, to %%@
our knowledge this is the first time that two second step transitions have been used for a laser ionization scheme. The spectroscopic %%@
capability of the combined laser system enables a maximal wavelength coverage representing a unique tool for on-line laser ion source %%@
facilities.
\subsection{Ion guide laser ionization of filament-produced yttrium atoms - March 2006} % 3.2.
\label{sec:06}
Under off-line conditions yttrium laser ions are formed through the resonant excitation and ionization process of atoms evaporated from an %%@
yttrium filament, as described in section \ref{sec:lisol}. The ions are evacuated from the laser ion guide, transported through a radio %%@
frequency sextupole device and injected into the mass separator via stages of differential pumping. After acceleration to 30 kV the beam is %%@
mass separated by a dipole magnet, allowing separation of nuclei and contaminants with a typical mass resolving power of the order of %%@
300-600 depending on the operational parameters of the guide and the front-end of the separator. Finally, the ions are detected on a set of %%@
multi-channel plates, downstream from the separator focal plane.
During transport through the ion guide there are many loss mechanisms for ions produced by laser ionization. As the atom or ion moves %%@
through the guide towards the exit nozzle it collides many times with buffer gas atoms and impurity molecules. This can result in chemical %%@
bond formation or trapping in a metastable state which both lead to losses as the laser scheme excites only from the atomic ground state.
\par
 The main loss mechanism in off-line conditions is the formation of molecules via reactions between an atom X or ion X$^+$ with a ligand %%@
molecule M in the presence of the buffer gas (in this case helium),
\begin{align} 
 \text{X}^{(+)} + \text{M}       & \leftrightarrow  \text{X}^{(+)}\text{M}^*   	\label{eq:4}	\\			 \text{X}^{(+)}\text{M}^* + %%@
\text{He} &  \rightarrow  \text{XM}^{(+)} + \text{He}     \; .      \label{eq:5}
\end{align}
The time evolution of the process of atoms or ions converting to molecules may be described by a reaction rate coefficient $k$ %%@
(cm$^3$s$^{-1}$) and the rate equation
\begin{align} 
\frac{dn}{dt}=-n k\left[M\right]
\label{eq:6}
\end{align}
where $n$ is the number of atoms or ions, and $\left[M\right]$ the ligand molecular concentration. A corresponding time constant for the %%@
formation of the molecular ion can be defined as
\begin{align} 
\tau=1/k\left[M\right]
\label{eq:7}
\end{align}
It is known that yttrium ions have a strong affinity towards binding with oxygen \cite{22koyanagi}. The main residual impurity in the %%@
buffer gas is water therefore the formation of YO$^+$ can either happen directly or via the dehydrogenation reaction
\begin{align} 
 \text{Y}^+ + \text{H-O-H} \rightarrow \text{H-Y}^+\text{-O-H} \rightarrow \text{H}_2\text{Y-O}^+ \rightarrow  \text{YO}^+ + \text{H}_2      %%@
\; .	     
\label{eq:8}
\end{align}
This has been discussed more thoroughly in \cite{23yuri} for the case of titanium. Within the literature one may find a reaction rate %%@
constant $k = 4.1 \cdot 10^{-10}$ cm$^3$s$^{-1}$ \cite{22koyanagi} for the reaction 
\begin{align} 
\text{Y}^+ + \text{O}_2 \rightarrow  \text{YO}^+ + \text{O} \; . 						
\label{eq:9}
\end{align}
With an impurity level of 1 ppm, which corresponds to a concentration $\left[M\right]$ of $\approx 4 \cdot 10^{12}$ atoms/cm$^3$ at a %%@
helium gas pressure of 150 mbar, the reaction time  $\tau$ for the formation of molecular YO$^+$ is 612 $\mu$s. This is a very short time %%@
in comparison to the bulk ion guide evacuation time of 480 ms and therefore it is imperative to study the effects of impurities on the %%@
yttrium laser ions and to have control over the gas purity to a level close to a ppb.
\par
\begin{figure}
\centering
\includegraphics[width=.95\linewidth]{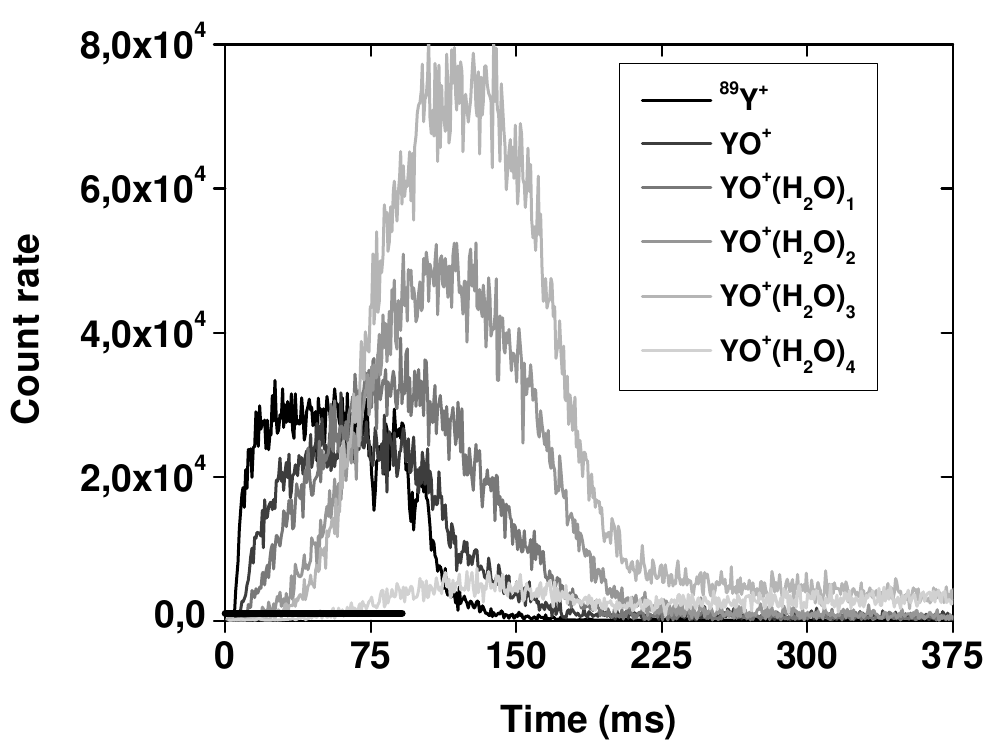}
\caption{ Time-of-flight profiles of yttrium laser ions and corresponding molecules. The black horizontal bar along the time axis %%@
represents the period the lasers were on.}
\label{fig:tofprof06}
\end{figure}

In this work, high-purity helium gas (grade 6.0, 99.9999\%) supplied by Linde Gas AG, Germany, has been used. The tubes of the gas feeding %%@
lines are made of stainless steel and were baked to $\approx$ 150 °C. The gas passes through a cold trap made of activated carbon, cooled %%@
with liquid nitrogen, to remove impurities. Additional purification was provided by a getter-based purifier (Saes MonoTorr PS4-MT15-R-2). %%@
According to \cite{23yuri} these cleaning steps should take the gas down to the sub-ppb impurity level. The yttrium filament was %%@
continuously heated throughout the measurements and therefore atoms were continually available for laser ionization during the laser ``on'' %%@
period. A helium pressure of 150 mbar was used throughout the off-line studies. The ion signal from the channel plates was fed into a %%@
multi-channel analyzer (MCA), triggered by the JYFLTRAP labview control program \cite{24hakala} in order to study the evacuation time of %%@
the ions and corresponding molecules, with a time resolution of 655.36 $\mu$s per bin. The first evacuation time profiles showing the %%@
evolution of the yttrium laser ions and molecular formation were taken in March 2006, and are shown in Fig. \ref{fig:tofprof06}.
% Fig. 11.
\par
\begin{figure}
\centering
\includegraphics[width=.95\linewidth]{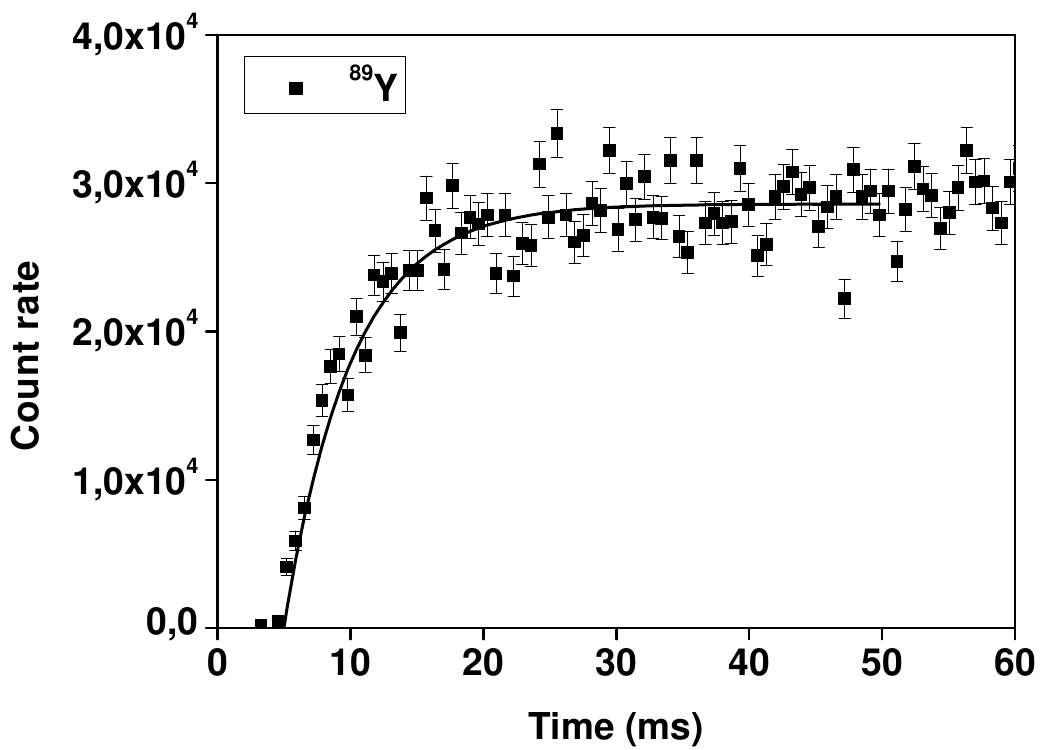}
\caption{ An exponential fit of the count rate towards saturation of laser ionized $^{89}$Y$^+$.
}
\label{fig:yorise}
\end{figure}

The laser radiation was pulsed on for the first 90 milliseconds of a full cycle of about four seconds. This was accomplished using the %%@
``single-arm'' mechanical shutter device discussed in section \ref{sec:shutter}, triggered by a delayed TTL signal from the JYFLTRAP %%@
control program. A delay of $\approx$ 5 ms occurs before the first ions were detected. This was caused by a time delay in the shutter and %%@
is not related to the time-of-flight of the ions. It can be seen from Fig. \ref{fig:tofprof06} that the laser-produced Y$^+$ ions reach a %%@
saturation level before the lasers are turned off. The structure seen in the falling edge of the yttrium time profile is related to a %%@
``ringing'' of the shutter arm which was addressed in the March 2007 data (section \ref{sec:07}) with the addition of the double-shutter %%@
device. The characteristic time to reach saturation can be related to a combination of the survival time of Y$^+$  via  Eq. \eqref{eq:7} %%@
against losses due to molecular formation and evacuation from the ion guide. It should be noted that the ion guide had a constant source of %%@
yttrium atoms produced from the heated filament prior to the moment of laser ionization and this level is in saturation with both %%@
evacuation from the guide and losses due to neutral atom-molecule formation. Fig.  \ref{fig:yorise} shows an exponential growth curve %%@
fitted to the $^{89}$Y$^+$ data resulting in a value of $\tau= 5.1(4)$  ms. 
As the evacuation time from the ion guide as a whole is far longer than this timescale, this saturation time is indicative of the survival %%@
time of yttrium ions during evacuation against molecular formation. Using the known conductance of the exit hole (112 cm$^3$s$^{-1}$) this %%@
saturation time translates into a corresponding effective volume for laser ionization of $\approx $ 0.57 cm$^3$. The laser ions which %%@
survive molecular formation are created close to the exit hole in a volume only 28\% of the total volume of the ionization channel of the %%@
LISOL ion guide (Fig. \ref{fig:lisol}). Ions created deeper within the gas cell have a high probability of forming molecules and thus %%@
reduce the Y$^+$ signal. 
\par
It can be seen from Fig. \ref{fig:tofprof06} that the formation of YO$^+$ ions is delayed with respect to the Y$^+$ ions, and each %%@
subsequent addition of a hydrate to the YO$^+$ molecule is delayed even further. The time behaviour indicates the following sequence of %%@
hydration of YO$^+$ ions:
\begin{align} 
 \text{YO}^+ + \text{H}_2\text{O}   &  \rightarrow  \text{YO}^+\text{(H}_2\text{O)}	 & 	(A &=123) \label{eq:10}	\\		     		  
\text{YO}^+\text{(H}_2\text{O)} + \text{H}_2\text{O} &  \rightarrow  \text{YO}^+\text{(H}_2\text{O)}_2	 &   (A&=141)   \label{eq:11} \\
\text{YO}^+\text{(H}_2\text{O)}_2 + \text{H}_2\text{O} &  \rightarrow  \text{YO}^+\text{(H}_2\text{O)}_3	&   (A&=159)   \label{eq:12} \\
\text{YO}^+\text{(H}_2\text{O)}_3 + \text{H}_2\text{O} &  \rightarrow  \text{YO}^+\text{(H}_2\text{O)}_4	& (A&=177)  \; .   %%@
\label{eq:13}  
\end{align}
The relative intensities, the saturation time of the Y$^+$ signal and the delay times of the subsequent molecules illustrated in Fig. %%@
\ref{fig:tofprof06} can provide information about specific impurity concentrations in an environment where the impurity level can be %%@
controlled. Although this is not the case in this work, the saturation time extracted from the experimental results can be combined with %%@
Eq. \eqref{eq:7} to estimate an impurity level. Under the assumption that the time scale for the dehydrogenation reaction of Eq. %%@
\eqref{eq:8} is similar to that of Eq. \eqref{eq:9} then, with an ion guide pressure of 150 mbar helium, the impurity level is $\approx$ %%@
0.1 ppm. This result was rather surprising due to the careful treatment of the gas purity, suggesting that the system was not as clean as %%@
we had expected. The bare atomic fraction of yttrium amounted to 10\% of the sum of all the species shown in Fig. \ref{fig:tofprof06}. This %%@
figure can be compared to Fig. \ref{fig:sim06} of \cite{23yuri} in which the gas was purified to the expected ppb level. In this figure the %%@
Ti$^+$ signal was scaled down by a factor of 50 to show it on the same scale as the subsequent molecules. The reaction coefficient $k$ for %%@
Ti$^+$ with oxygen is $4.6 \cdot 10^{-10}$ cm$^3$s$^{-1}$ \cite{22koyanagi} which is the same order of magnitude as for yttrium. 
\par
It is of interest to get a deeper understanding of the time scales of formation of individual molecules from the experimental data. One %%@
important difference between the laser system used in this work and that of the LISOL system is the repetition rate of the lasers. It is %%@
possible to study the evolution of laser ions created in the LISOL system from a single laser shot, i.e. at a repetition rate of %%@
effectively 1 Hz, in which the time between laser pulses is longer than the evacuation time of the cell. In this case a fresh portion of %%@
atoms is available before the next pulse arrives. In this work a high repetition rate laser system is used. The double-shutter mechanism %%@
discussed in section \ref{sec:shutter} was used to reduce the 10 kHz laser system to 2-3 individual laser pulses, each 30-50 ns wide and %%@
separated by 100 $\mu$s. However, due to the reduced energy per pulse of the high repetition rate system the ionization efficiency was most %%@
likely too low to detect any laser ions and therefore a different treatment of the data is needed in order to study the molecular formation %%@
on single shot time scales.
\par
\begin{figure}
\centering
\includegraphics[width=.95\linewidth]{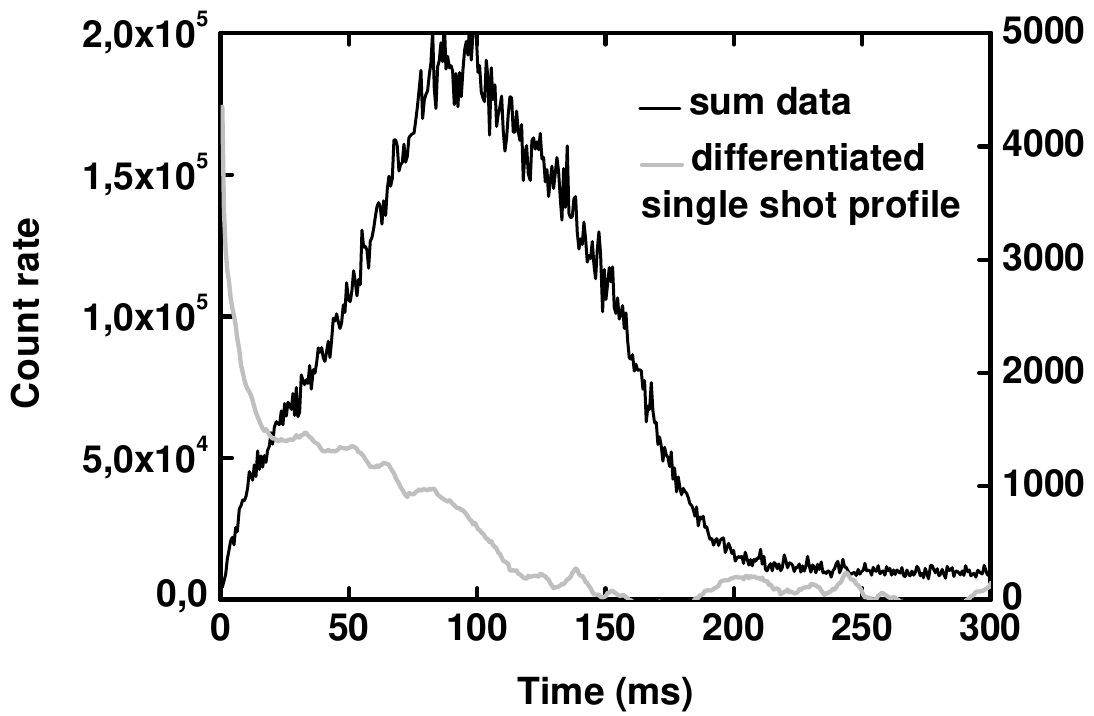}
\caption{ The summed data from Fig. \ref{fig:tofprof06} and the corresponding differentiated single shot evacuation time profile, %%@
calculated according to  Eq. \eqref{eq:16} with $t_{off}$ = 90 ms. }
\label{fig:sumdata06}
\end{figure}
\begin{figure}
\centering
\includegraphics[width=.95\linewidth]{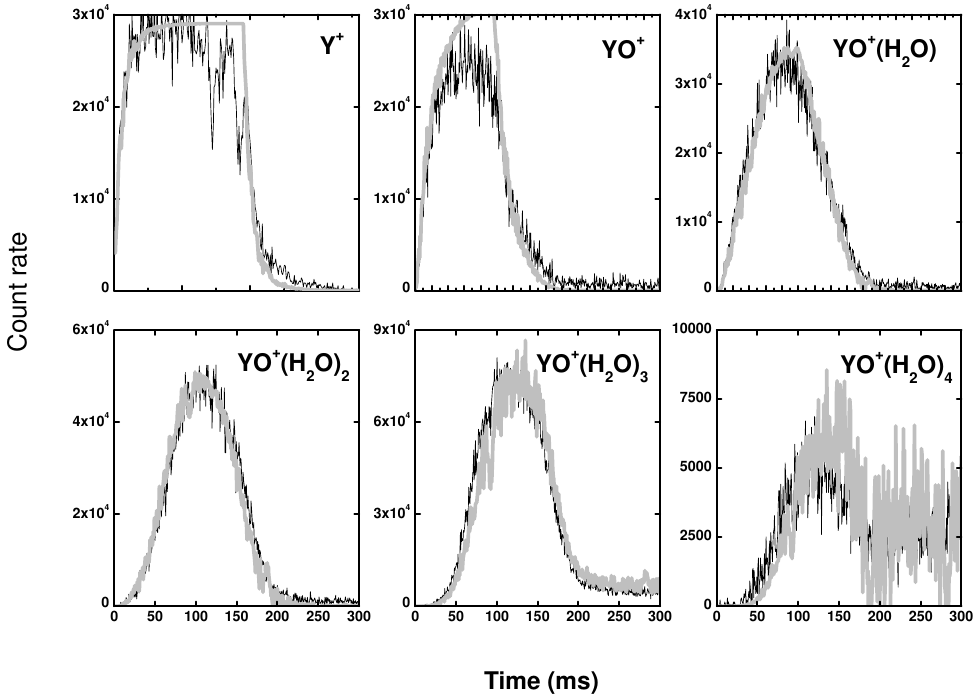}
\caption{ Comparison between the simulated single-shot profiles (grey line) and the experimental data (black line). }
\label{fig:sim06}
\end{figure}

The reaction rate $k\left[M\right]$ reflects the probability that a single ion will be converted into a molecule during its evacuation from %%@
the ion guide. In a 10 kHz laser system each shot provides a new sample of yttrium ions which has a probability of molecular formation. The %%@
time-of-flight profiles seen in Fig. \ref{fig:tofprof06} are therefore a convolution of all single-shot events, refreshed with every %%@
subsequent laser shot. The procedure to obtain the molecular reaction rates from the data is performed by applying the following four %%@
steps:
\begin{itemize}
\item Data from Fig. \ref{fig:tofprof06} is summed up to obtain an overall Y$^+$ signal. In this manner the sum can be thought of as the %%@
total current extracted from the ion guide before mass separation.
\item   A single laser shot ion evacuation profile is extracted from the summed data by taking the derivative of the overall time %%@
distribution.
\item The molecular formation process is applied to the yttrium ions evacuated from a single shot.
\item The single-shot data for the different molecular compounds is reintegrated to the summed data and compared with the experimental %%@
data.
\end{itemize}
In this approach it is assumed that the molecular formation process is a closed system and losses due to, for example, diffusion and %%@
neutralization with electrons caused by the passage of the laser beam are not taken into account in this simple model. One can therefore %%@
write
\begin{multline} 
\text{Y}^+ +\text{YO}^+ +\text{YO}^+\text{(H}_2\text{O)}+\text{YO}^+\text{(H}_2\text{O)}_2 \\ 
+\text{YO}^+\text{(H}_2\text{O)}_3+\text{YO}^+\text{(H}_2\text{O)}_4=\text{const.}	
\label{eq:14}
\end{multline}
According to this assumption the total sum of the data represents the Y$^+$  signal after evacuation from the ion guide without any %%@
molecular formation. In the second step a relationship between the summed signal, $total(t)$, and a single laser shot, $shot(t)$, is %%@
established. It is assumed that the atomic yttrium vapour recovers during the time $t$ between two consecutive laser shots. This is a %%@
reasonable conclusion for two reasons: firstly, the filament is heated continuously and, secondly, the product of ionization efficiency for %%@
an individual laser shot with the number of laser shots that the irradiated volume sees during the evacuation time is $<<$ 1. The total %%@
signal is therefore the integral of individual shots shifted in time
\begin{align} 
total(t)=\int_0^{t_{off}} shot(t-s)\, ds
%  .								     (15)
\label{eq:15}
\end{align}
where $t_{off}$ denotes the time when the laser radiation is turned off. This is the case in which the laser duty cycle is much smaller %%@
than the response time of the system. A single laser shot evacuation profile can be reproduced by taking the derivative of the total signal %%@
\eqref{eq:15}, 
\begin{align}
\begin{split} 
\frac{d}{dt}total(t)&=\frac{d}{dt} \int_0^{t_{off}} shot(t-s)\, ds \\
& =shot(t)-shot(t-t_{off})  \\
\rightarrow shot(t) &=\frac{d}{dt} total(t)+shot(t-t_{off}) \, .
\end{split}
\label{eq:16}
\end{align}
Note that the last expression in the formula, $shot(t-t_{off})$, gives zero for $t<t_{off}$ as $shot(t<0)=0$. Therefore for $t<t_{off}$ the %%@
single shot distribution is simply the derivative of the total signal $total(t)$, while for $t>t_{off}$ the contribution of %%@
$shot(t-t_{off})$ has to be taken into account.
Fig. \ref{fig:sumdata06} illustrates the summed time-of-flight profile of yttrium and the molecular compounds individually shown in Fig. %%@
\ref{fig:tofprof06}, and the corresponding derivative of the sum signal.
% Fig. 13.
 Note that the sum data starts at $t=0$ seconds, unlike the data in Fig. \ref{fig:tofprof06} which has an initial delay of  $\approx$ 5 ms. %%@
In Fig. \ref{fig:sumdata06} the delay has been removed as it does not originate from the gas cell evacuation. The single shot time profile %%@
was processed with an adjacent averaging smoothing filter taking into account the closest 100 neighbouring points. The sum profile clearly %%@
has some structure within it which is directly related to the intensity balance of the individual molecules and is reflected in the %%@
differentiated signal which can be used to illustrate the evacuation time profile of a single laser shot. The single shot profile is not a %%@
smooth distribution in time rather there is a fast decay in the first few milliseconds followed by a long tail, extending beyond 300 ms, %%@
which is not surprising as the total evacuation time of the ion guide based on the volume and conductance of the exit hole is 480 ms. 
\par
Assuming that the molecular formation follows the evolution process of Eq. (\ref{eq:10}-\ref{eq:13}) one can use  Eq. \eqref{eq:6} to %%@
formulate a set of rate equations of the form
\begin{align}
\begin{split} 
\frac{d \left[Y^+\right]}{dt}        &=       -k_1     \left[M\right]      \left[Y^+\right] \\
\frac{d \left[YO^+\right]}{dt}        &=      k_1     \left[M\right]      \left[Y^+\right]-
k_2     \left[M' \right]      \left[YO^+\right]  \\
\frac{d \left[YO^+(H_2O)\right]}{dt}        &=      k_2   \left[M'\right]      \left[YO^+\right]-k_3    \left[M' \right]      %%@
\left[YO^+(H_2O)\right]   \\
\frac{d \left[YO^+(H_2O)_2\right]}{dt}        &=      k_3   \left[M'\right]      \left[YO^+(H_2O)\right] \\
& \qquad -k_4    \left[M' \right]      \left[YO^+(H_2O)_2\right]   \\
\frac{d \left[YO^+(H_2O)_3\right]}{dt}        &=      k_4   \left[M'\right]      \left[YO^+(H_2O)_2\right] \\
 & \qquad -k_5    \left[M' \right]      \left[YO^+(H_2O)_3\right]   \\
\frac{d \left[YO^+(H_2O)_4\right]}{dt}        &=      k_5   \left[M'\right]      \left[YO^+(H_2O)_3\right] 
\label{eq:17}
\end{split}
\end{align}
where $\left[M\right]$ is the sum of the impurity water and oxygen atom concentrations, and $\left[ M'\right]$ is the water impurity %%@
concentration alone. The reaction rate coefficient $k_1$ is known from literature \cite{22koyanagi}, while $k_2$ to $k_5$ are rate %%@
coefficients  for the subsequent molecular formation reactions. These are expected to be within the same order of magnitude as $k_1$ %%@
\cite{25bohme}. In this closed system of rate equations in which the condition of total current conservation is fulfilled (Eq. %%@
\eqref{eq:14}) the initial starting condition is that $\left[Y^+\right](t=0) = 1$. The five reaction rates $k_n\left[M \, \text{or}\,  %%@
M'\right]$ form the free parameters within this model.
\par
\begin{figure}
\centering
\includegraphics[width=.95\linewidth]{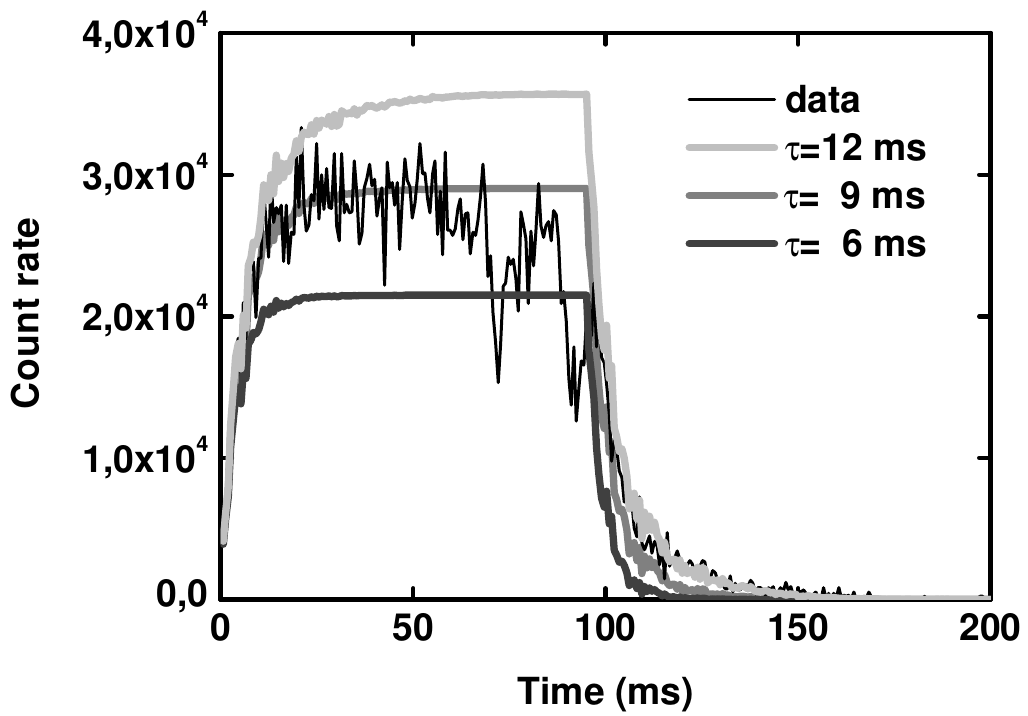}
\caption{ The sensitivity of the simulated time profile for $\left[Y^+\right]$ as a function of different reaction rate parameters %%@
$k_1\left[M\right]$. }
\label{fig:sensitivity06}
\end{figure}

A single yttrium atom in the ion guide is ionized by the laser at $t=0$. The probability for molecular formation can be determined when the %%@
ion has finally reached the exit hole and is evacuated. Applying the rate equation model of   Eq. \eqref{eq:17} to the differentiated %%@
single shot profile of Fig. \ref{fig:sumdata06} results in a typical single shot evacuation time profile for each ionic compound in the gas %%@
cell. These individual profiles are then reintegrated using  Eq. \eqref{eq:15}. The five reaction rate parameters are adjusted to optimize %%@
the agreement of the simulation to the ratios of the different molecules observed in the experimental data. Fig. \ref{fig:sim06} shows the %%@
individual experimental time-of-flight profiles of Fig. \ref{fig:tofprof06} compared with the reintegrated single shot profiles. 
% Fig.14. 
In order to illustrate the sensitivity of the reaction rate parameter to the comparison with experimental data, Fig. %%@
\ref{fig:sensitivity06} shows how the simulated time profile of $\left[Y^+\right]$ varies as a function of $k_1\left[M\right]$. 
% Fig.15. 
Recall that the model should reproduce the signal amplitudes as the total sum is conserved. The relative error in the fitting procedure is %%@
approximated to 10\%. The time constants for molecular formation extracted from the reaction rate parameters that provide the optimum fits %%@
in Fig. \ref{fig:sim06} are given in table \ref{tab:sim06}.
% Table 4
\begin{table}
\centering
\caption{ The molecular formation time (ms) extracted from the fitted reaction rates.}
\begin{tabular}{ccccc}
\hline
$\tau_1=$	   &
$\tau_2=$	   &
$\tau_3=$	   &
$\tau_4=$	   &
$\tau_5=$	   \\ 

$1/k_1\left[M\right]$	   &
$1/k_2\left[M'\right]$	   &
$1/k_3\left[M'\right]$	   &
$1/k_4\left[M'\right]$	   &
$1/k_5\left[M'\right]$	   \\ \hline

9	& 14	& 19	& 31     & 	400 \\ \hline
\end{tabular}
\label{tab:sim06}
\end{table}
\par
It can be seen from $\tau_1$ that the model overestimates the experimental saturation time of $\left[Y^+\right]$ (Fig. \ref{fig:yorise}) by %%@
$\approx$ 4 ms. The discrepancy appears to be related to the fast falling edge of the single shot profile rather than missing physics in %%@
the model. The long formation time $\tau_5$ is due to poor statistics and so had an uncertainty of 50\% associated with the fit. This %%@
fourth hydrate addition is laser-related and has a very long evacuation time profile, even though the count rate is considerably smaller %%@
than the other molecules (Fig. \ref{fig:tofprof06}). To a first approximation, the single shot treatment presented in this work appears to %%@
reproduce the experimental evacuation time profiles reasonably well. In ideal conditions this model, along with the experimental time %%@
profiles, could be used to extract individual impurity concentrations within the gas cell. The source of the impurities would not be an %%@
issue only the relative concentrations that are reflected in the experimental data. 
\par
In this work the problem to extract definite values for the concentrations of oxygen and water lies in the mechanism used to form YO$^+$, %%@
which is either through a direct reaction of yttrium with oxygen, or with water via Eq. \eqref{eq:8}. However, if we naively assume that %%@
YO$^+$ is formed directly from Eq.  \eqref{eq:9} then we can estimate an impurity level of oxygen in the ion guide using the results %%@
obtained in table 4. The number density of helium atoms in 150 mbar pressure is $3.98 \cdot 10^{18}$ atoms/cm$^3$. Using a value of 9 ms %%@
for $\tau_1$ the impurity concentration $\left[M\right]$ is equal to $2.71\cdot 10^{11}$ atoms/cm$^3$. Although earlier it was stressed %%@
that $\left[M\right]$ defined the sum of water and oxygen impurities, here we assume that as the reaction rate $k_1$ is that of  Eq. %%@
\eqref{eq:9} $\left[M\right]$ therefore refers to the level of oxygen. The subsequent impurity level is therefore 68 ppb assuming that the %%@
source of impurities is from the gas rather than the walls of the ion guide or other sources. Furthermore, a value of $k_2$ of $1.9 \cdot %%@
10^{-10}$ cm$^3$ s$^{-1}$ has been measured for the reaction $\text{YO}^+ + \text{D}_2\text{O} \rightarrow \text{YO}^+\text{(D}_2\text{O)}$ %%@
\cite{25bohme}. Using the value $\tau_2$ obtained in table \ref{tab:sim06} the impurity concentration $\left[M'\right]$ within the gas cell %%@
is calculated to be $3.76\cdot 10^{11}$ atoms/cm$^3$. Again, assuming that $\left[M'\right]$ is from the gas then the impurity of water is %%@
at a level of 94 ppb. These numbers can be compared to the impurity level of 0.1 ppm obtained from the exponential growth fitted to the %%@
yttrium data in Fig. \ref{fig:yorise}.
\par
The first set of measurements described here was taken under rather unfavourable conditions in March 2006. Experimentally only the rise %%@
time of yttrium could be compared to the molecular formation model as all other molecules did not reach a saturation level. With hindsight %%@
a more meaningful comparison of the reaction rates extracted from the single-shot time profiles with the experimental data could have been %%@
made provided all molecules reached saturation. Importantly, as discussed briefly earlier, the extracted impurity level of $\approx$ 0.1 %%@
ppm was two orders of magnitude higher than expected following the careful treatment of the gas purity. Indeed, following this work two %%@
problems were discovered. The first was a possible leak in the final part of the gas line going directly to the ion guide, while the second %%@
was a leak in the venting valve attached directly to the IGISOL chamber which led to a pressure of $\approx$ 10$^{-2}$  mbar (measured %%@
directly on the chamber) without helium gas flowing through the ion guide. Typically the pressure without buffer gas should be in the range %%@
of a few 10$^{-5}$ mbar to low 10$^{-4}$ mbar depending on how soon after opening the vacuum chamber the pressure is measured. The %%@
experiment was therefore repeated in March 2007 along with a study of the effect of introducing a leak into the vacuum chamber while the %%@
ion guide is running at typical operating pressures. This work will be discussed in the following section.
\subsection{Ion guide laser ionization of filament-produced yttrium atoms - March 2007} % 3.3.
\begin{figure}
\centering
\includegraphics[width=.95\linewidth]{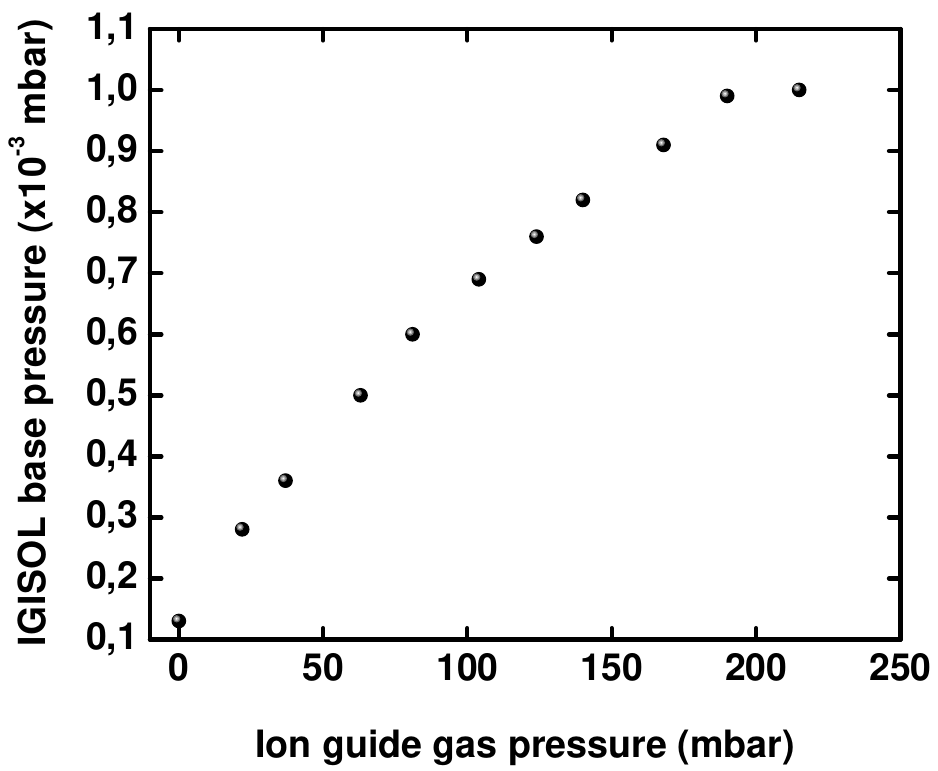}
\caption{ IGISOL vacuum chamber pressure as a function of ion guide pressure.
}
\label{fig:igisolpressure}
\end{figure}

\label{sec:07}
The same procedure as described in the previous section was followed in order to prepare the ion guide and to ensure clean gas purity %%@
conditions. At the beginning of the experiment a base line pressure of $1.3\cdot 10^{-4}$ mbar was measured in the IGISOL chamber without %%@
helium gas. These background conditions were approximately two orders of magnitude better than in the March 2006 experimental run. The %%@
chamber pressure was then monitored as a function of ion guide pressure, shown in Fig. \ref{fig:igisolpressure}.
% Fig.16. 
 At an operating pressure of 150 mbar, the chamber pressure was $\approx 8.4 \cdot 10^{-4}$ mbar. These numbers should not be taken as %%@
absolute values as the cold cathode gauge (model TPR 020) measurements have a dependence on the type of gas and it is calibrated for air. %%@
Additionally, the vacuum gauges are mounted on the opposite wall of the IGISOL chamber to that of the exit nozzle of the ion guide and %%@
therefore background measurements in the presence of flowing helium gas do not reflect the immediate surroundings of the expanding gas jet.
 \par
\begin{figure}
\centering
\includegraphics[width=.95\linewidth]{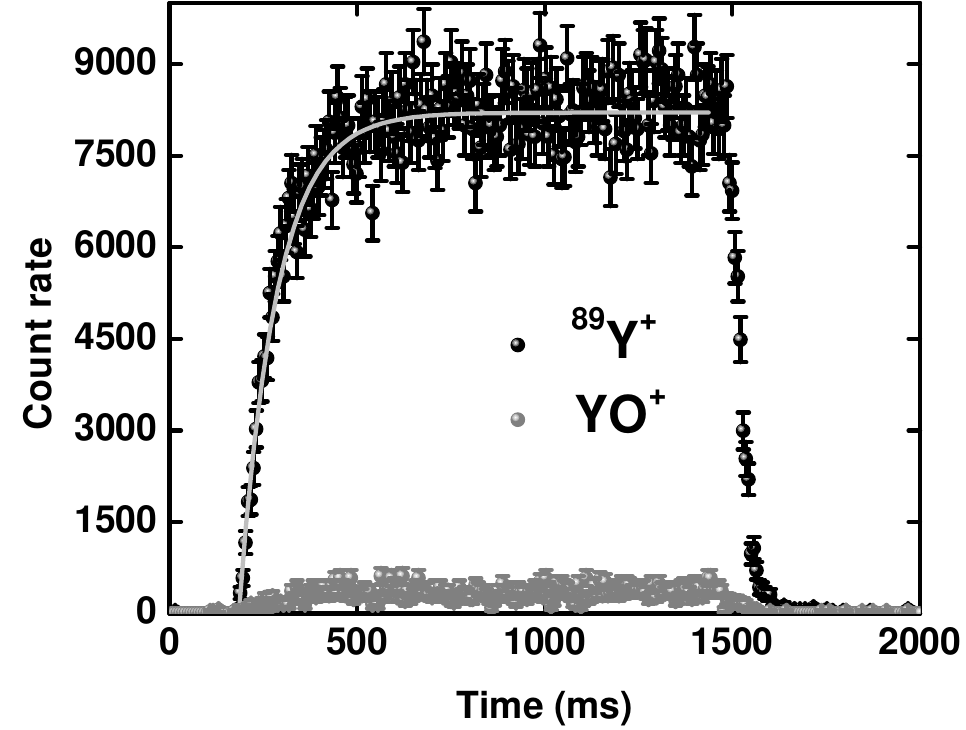}
\caption{ Evacuation time profiles of Y$^+$ and YO$^+$. The laser was on from 100 ms to 1.4 s. The grey line shows an exponential fit to %%@
the yttrium data. 
}
\label{fig:tofprof07}
\end{figure}

The double-arm mechanical shutter was used to ensure a better control over the laser arrival time and also to reduce any effects related to %%@
``ringing'' which is clearly visible on the falling slope of the yttrium time profile in Fig. \ref{fig:sensitivity06}. The laser was turned %%@
on after 100 ms and then at a time of 1.4 seconds was turned off, allowing a complete saturation of any molecules. The evacuation time %%@
profiles of the data taken under these new conditions are shown in Fig. \ref{fig:tofprof07}.
% Fig.17.
\par 
	The most striking difference between this data and that of Fig. \ref{fig:tofprof06} is that yttrium is now the dominant species over %%@
the next most abundant molecule extracted from the ion guide, yttrium oxide, by a factor of $\approx$ 20. No subsequent addition of %%@
hydrates could be observed at the expected mass numbers. Exponential fits to the rise time of the yttrium and the yttrium oxide data %%@
yielded values of 96(10) ms and 135(7) ms respectively. The errors on the fits arise from a deviation from an exponential behaviour at the %%@
beginning of the rising edge, and therefore are conservatively increased to account for this effect. The rise time for yttrium, 96(10) ms %%@
can be directly compared with the rise time fitted in the March 2006 data, 5.1(4) ms (Fig. \ref{fig:yorise}). With this new rise time and %%@
the ratio between yttrium and yttrium oxide, the primary effect on the time scale is now the evacuation time. By combining the rise time %%@
from this new data with the exit hole conductance, the effective volume for ion survival is $\approx$ 11 cm$^3$, approximately 20 times %%@
larger than in the earlier data. 
\par
One interesting detail seen in Fig. \ref{fig:tofprof07} is that the fall time of the yttrium signal is shorter than the rise time. This may %%@
be an indication of another effect that is not attributed to evacuation or molecular formation (which is minimal here). One possibility is %%@
a loss mechanism investigated and discussed in \cite{23yuri}, that of recombination in the presence of ion-electron pairs created by the %%@
laser ionization process. Under the assumption that there are no losses of the laser ions through diffusion and molecular formation within %%@
the ion guide, and transport through the mass separator, a lower limit of the density of ion-electron pairs can be estimated from the total %%@
count rate of $^{89}$Y seen in Fig. \ref{fig:tofprof07} and the ionization volume within the ion guide. With a total count rate of $\approx$ 8000 %%@
ions/s and a volume of 0.6 cm$^{3}$ (a cylinder of length 87 mm and diameter 3 mm) the ionization-density rate $Q\approx 10^4$ %%@
ion-electron pairs/cm$^3$s. The time scale $\tau$ for the production of ion-electron pairs reaching an equilibrium with three-body %%@
recombination is given by
	\begin{align}
 \tau=\frac{1}{\sqrt{Q \cdot \alpha}}
 	\end{align}
	where $\alpha$ is the pressure-dependent recombination coefficient. In 150 mbar helium gas, $\alpha$ is calculated to be $1.18 \cdot %%@
10^{-7}$ cm$^{3}$s$^{-1}$ \cite{alpha}. A time scale $\tau$ of $\approx$ 25 s can be deduced, indicating that the number of electrons is not likely to %%@
be high enough to significantly reduce the fall-time of the signal. Therefore the explanation of the effect is still subject to %%@
speculation.		
	
%	The single-shot model does not include such effects. However, we wish to stress that this is only speculation.
	\par
	In this instance it is not clear whether the rise time can be used to extract an impurity level as in the case of the March 2006 data. %%@
This is because the new data reveals that the evacuation is playing a more dominant role and therefore the effect of impurities need to be %%@
deconvoluted from the time profile. However, an upper limit can be determined in the same way as in section \ref{sec:06} and so with a %%@
helium pressure of 150 mbar in the ion guide, the impurity level can be estimated to be $\approx$ 6 ppb. This is a considerable improvement %%@
to the previous data and is more realistic based on the gas purity control discussed in detail in \cite{23yuri}.
	\par
\begin{figure}
\centering
\begin{minipage}[t]{0.49 \linewidth}
\includegraphics[width=.95\linewidth]{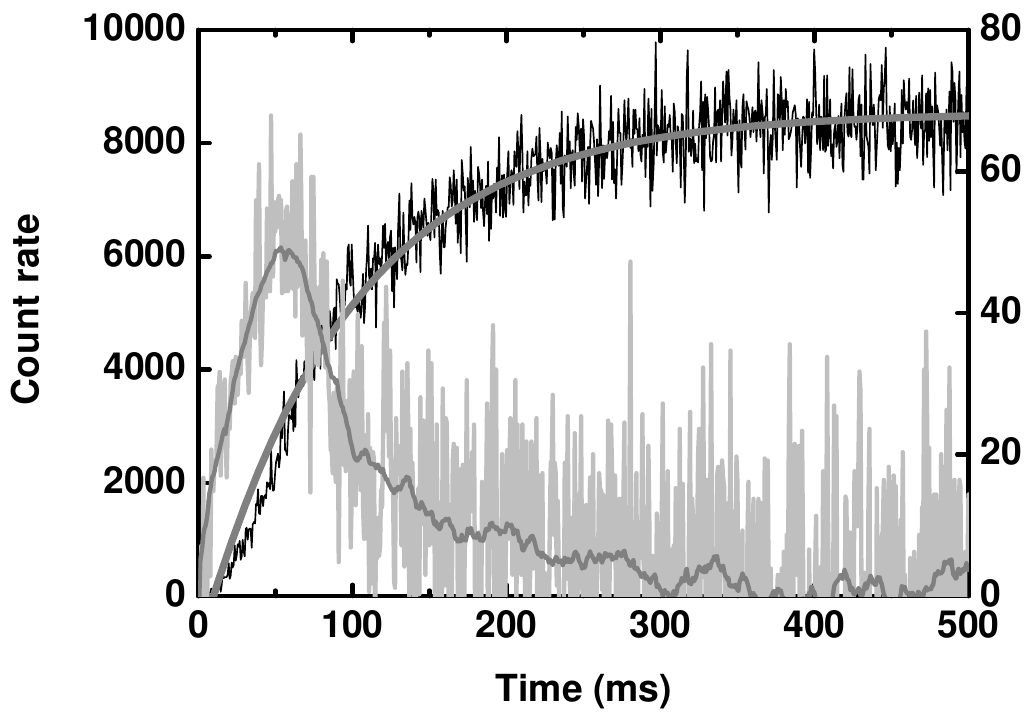}
\end{minipage} \hfill
\begin{minipage}[t]{0.49 \linewidth} 
\includegraphics[width=0.95 \linewidth]{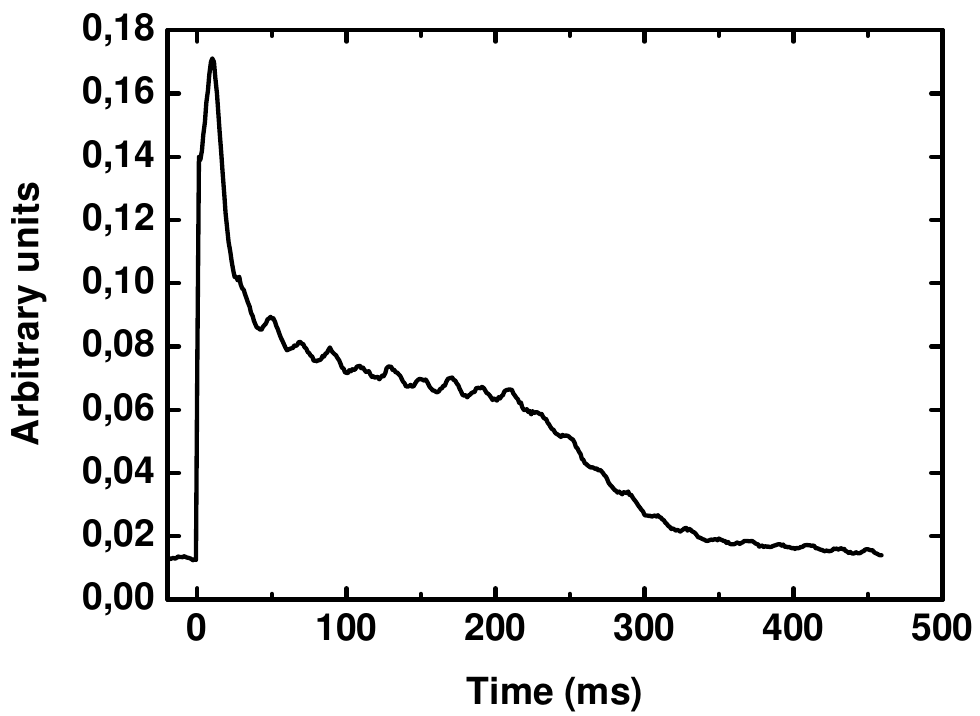} 
\end{minipage}
 \caption{ Summed data from Fig. \ref{fig:tofprof07} and the corresponding differentiated single shot evacuation time profile (left). An %%@
exponential fit to the summed profile is shown  and an adjacent averaging technique using 50 bins has been applied to the differentiated %%@
data in order to smooth out the fluctuations. On the right a true single laser shot evacuation time profile of ions produced by laser %%@
ionization with the LISOL low-repetition rate laser system is illustrated for comparison.
}
\label{fig:sumdata07}
\end{figure}

	The same procedure as in section \ref{sec:06} is then applied to extract a single laser shot evacuation profile from the summed data. %%@
The saturation of the data in this case means that the single shot can be derived simply from the derivative of the summed data according %%@
to Eq. \eqref{eq:16}. The simulation only takes into account the rising edge of the experimental data leading to saturation, corresponding %%@
to a length of the single shot of 500 ms. Following the differentiation of the summed signal and applying the rate equations for the known %%@
molecular formation the single shot profiles are reintegrated and compared with the experimental evacuation profiles. The summed %%@
experimental data and the differentiated signal are shown on the left in Fig. \ref{fig:sumdata07}. It is clear that as the yttrium signal %%@
dominates, the summed profile closely follows that of the yttrium profile unlike the experimental data in section \ref{sec:06}. An %%@
exponential fit to the data yields a rise time of 99(10) ms, in agreement with the yttrium rise time of 96(10) ms. The saturation of the %%@
data is clearly helpful in extracting a reliable single shot profile which is seen to peak at approximately 50 ms before decreasing %%@
steadily to longer evacuation times. The peak of the rise time corresponds well with gas flow simulations that show the atomic vapour %%@
density is highest where the gas flow starts to converge towards the ionization channel of the ion guide.
\par
	For comparison, a true single shot evacuation time profile obtained using the low repetition rate laser system at the LISOL facility is %%@
shown on the right in Fig. \ref{fig:sumdata07} \cite{27yuri}. 
% fig18
The data was taken using the same ion guide geometry and buffer gas as in the present work, using the lasers to ionize nickel atoms from a %%@
heated filament. In Fig. \ref{fig:sumdata07} the fast rise of the LISOL single shot profile, peaking at approximately 10 ms, arises from %%@
the evacuation of the ionization channel. The long tail of the distribution can be explained by the evacuation of the main body of the ion %%@
guide. While there is a rather good agreement between the two single shot evacuation profiles in terms of the time scale of the full %%@
evacuation of the ion guide ($\approx$ 400 ms) the shapes of the profiles look rather different. A probable reason can be found in the %%@
difference between the laser-atom spatial overlap. In the LISOL system the laser beams are unfocussed and the laser fills the ionization %%@
channel (10 mm diameter). In the case of the IGISOL system the lasers enter the ion guide and are focused to a beam waist of $\approx$ 3 mm %%@
at the nozzle in order to provide sufficient laser intensity to saturate the required atomic transitions. The single shot profile in the %%@
latter case reflects the situation in which the lasers are not only probing a smaller volume of the ion guide but are more sensitive to any %%@
misalignment of the laser beam on the symmetry axis, which translates into a higher sensitivity to different regions of the gas flow. More %%@
experiments have to be performed to study the dependence of the evacuation profile on the geometry of the ion guide, the laser ionization %%@
region and filament position in detail. 
\par
\begin{figure}
\centering
\begin{minipage}[t]{0.49 \linewidth}
\includegraphics[width=.95\linewidth]{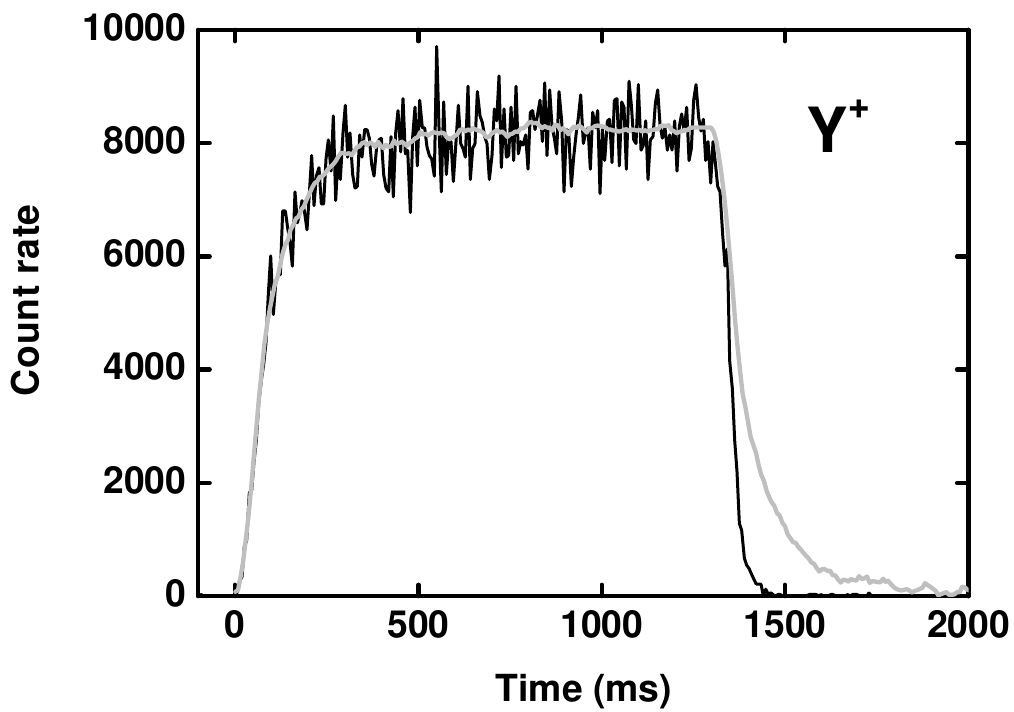}
\end{minipage} \hfill
\begin{minipage}[t]{0.49 \linewidth} 
\includegraphics[width=0.95 \linewidth]{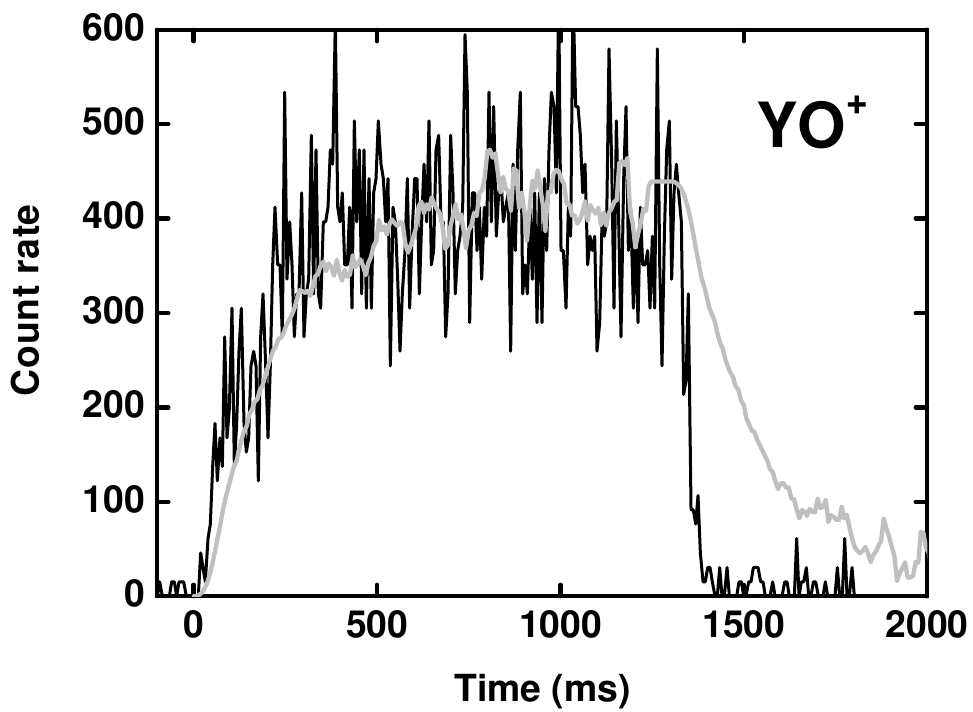} 
\end{minipage}
\caption{Comparison between the simulated single shot profiles (grey line) and the experimental data (black line) using an optimum $\tau_1$ %%@
of 2.3 seconds for the differentiated single shot.
}
\label{fig:sim07}
\end{figure}

	After applying the rate equation model of  Eq. \eqref{eq:17}, Fig. \ref{fig:sim07} shows the individual ionic time profiles of the %%@
yttrium and yttrium oxide after evacuation from the gas cell.
% fig19
	 In this simulation the higher order reaction rate coefficients are assumed to be zero, reflecting the experimental data. The simulated %%@
time profiles have been created using an optimum $\tau_1$ of 2.3 seconds. This fit parameter can be compared to $\tau_1$ from table %%@
\ref{tab:sim06} (9 ms) and reflects the fact that the evacuation time completely dominates the saturation profile. The ratio of yttrium to %%@
yttrium oxide has a very strong dependence on the value of $\tau_1$. One may question why the experimentally fitted rise time of 96(10) ms %%@
is so much shorter than 2.3 seconds, and whether the simulation could reproduce the data with a shorter $\tau_1$. The evacuation time, %%@
however, is not the same as the optimized value of the reaction rate coefficient used to correctly fit the ratios of Y$^+$ to YO$^+$. This %%@
observation illustrates that the single shot model is a powerful tool to directly probe the level of impurities in a system dominated by %%@
the evacuation time, where simply fitting the rise time of the atomic species can only provide an upper limit. Therefore, following the %%@
same assumptions used in section \ref{sec:06} an oxygen impurity concentration $\left[M\right]$ of $1.06 \cdot 10^9$ atoms/cm$^3$ can be %%@
extracted, which can be translated into an impurity level of 0.27 ppb if the source of the impurity is in the helium gas. This is a factor %%@
of 20 lower than the upper limit estimated from fitting the rise time of the data.
	\par
	In Fig. \ref{fig:sim07}  the simulation clearly reproduces the rising edge and amplitude of the time profiles, however unlike Fig. %%@
\ref{fig:sim06} a striking deviation exists between the experimental falling edges and that of the simulation, which overestimates the %%@
decay time. This feature cannot be explained by the model as by definition the slope of the rising and falling edges is dominated by the %%@
time scale of the single shot event, and therefore should be the same. As mentioned previously, the reason for this discrepancy is not %%@
fully understood. 
\par
\begin{figure}
\centering
\begin{minipage}[t]{0.49 \linewidth}
\includegraphics[width=.95\linewidth]{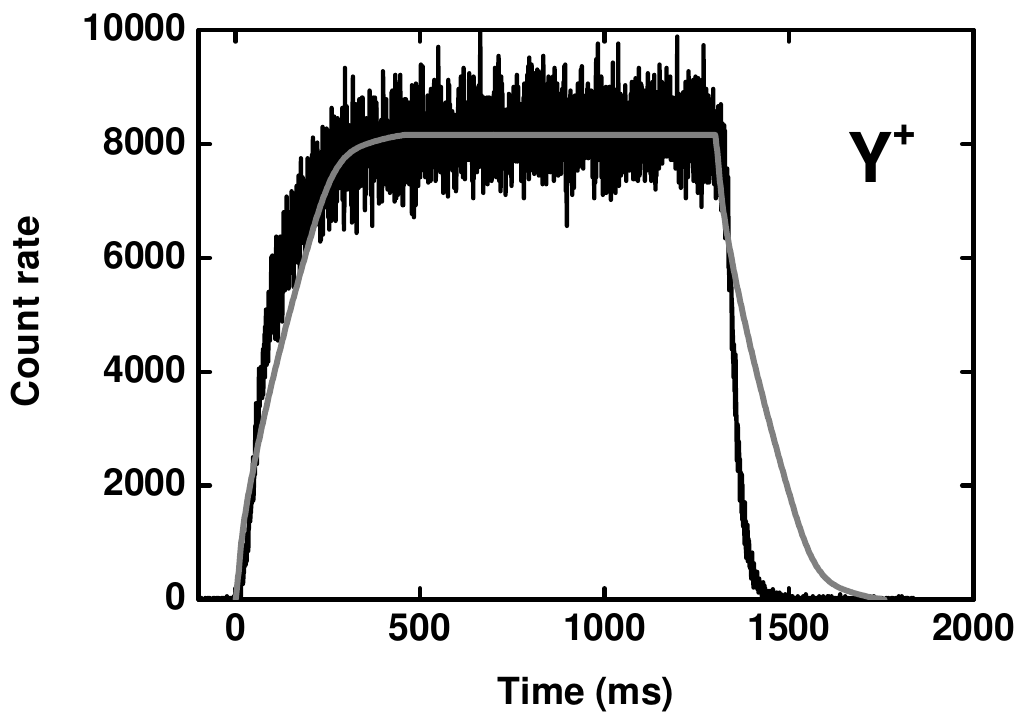}
\end{minipage} \hfill
\begin{minipage}[t]{0.49 \linewidth} 
\includegraphics[width=0.95 \linewidth]{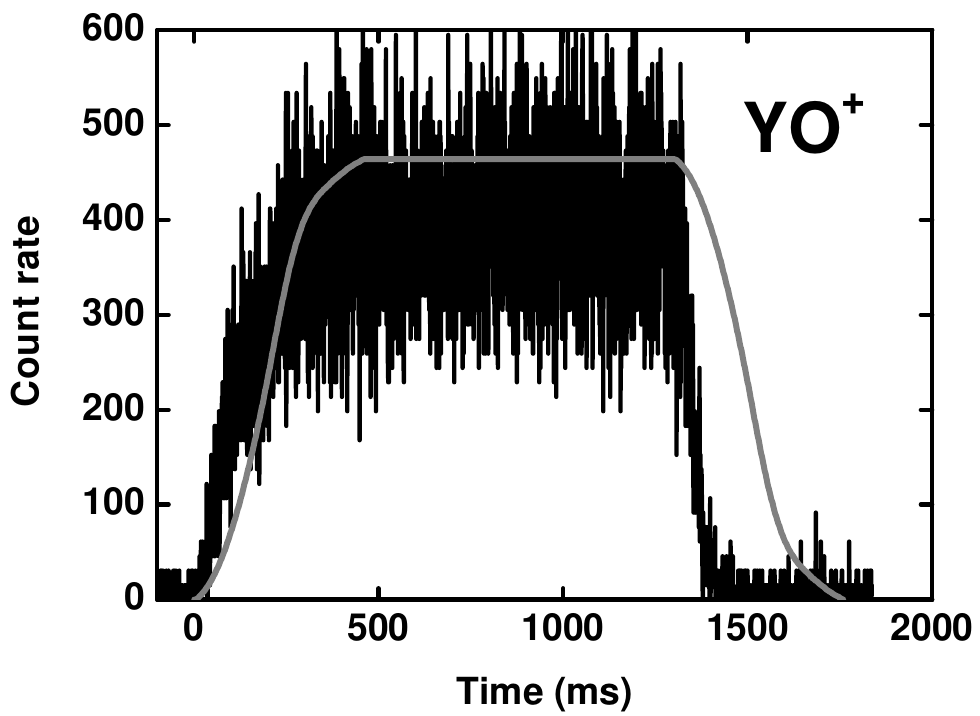} 
\end{minipage}
\caption{ Comparison between the simulated LISOL single shot profiles (grey line) and the experimental data (black line) using a $\tau_1$ %%@
of 2.3 seconds.}
\label{fig:sim07lisol}
\end{figure}

	Finally the sensitivity of the model towards the shape of the single shot evacuation profile was investigated. The experimentally %%@
determined laser shot profile from the LISOL data, illustrated in Fig. \ref{fig:sumdata07}, was used as an input for the simulation taking %%@
the same time scale $\tau_1$ for molecular formation (2.3 s) extracted from the optimum fit of the differential profile to the yttrium %%@
experimental data. The results using the LISOL single shot are shown in Fig. \ref{fig:sim07lisol}.
% fig 20
	 A constant scaling factor for both Y$^+$ and YO$^+$ was applied to take into account the arbitrary $y$-axis scale of the LISOL single %%@
shot profile. The simulated ratio of Y$^+$ to YO$^+$ is affected only slightly by the change of the profile, however compared with the fits %%@
in Fig. \ref{fig:sim07} it is noticeable that the rising edge of the data is not as well described. The difference in the simulated rising %%@
edges simply reflects the different shapes of the two single shot profiles, and in turn the differing laser-atom overlap geometries and %%@
thus laser ion evacuation times. Unsurprisingly, the choice of profile does not solve the discrepancy between the simulated falling edge %%@
and the experimental data.
\par
In order to understand the importance of the baseline vacuum chamber pressure in a controlled fashion, a needle valve was attached directly %%@
to the IGISOL vacuum chamber. The motivation for this study arose from the poor conditions experienced in the 2006 experiment. The work of %%@
the LISOL group has shown that the vacuum chamber can be opened with no detrimental effects to the purity conditions as long as a gas flow %%@
is maintained within the ion guide \cite{27yuri}. During the following tests it is important to note that a helium gas pressure of 150 mbar %%@
was maintained in the ion guide throughout the measurements, and the IGISOL vacuum chamber pressure was measured at this gas pressure. %%@
Initially, when the needle valve was closed, the yttrium and yttrium oxide ion count rate was monitored starting from high purity %%@
conditions, and then subsequently worse conditions as the getter was bypassed, the liquid nitrogen cooled cold trap was bypassed and the %%@
helium gas bottle was changed from a grade 6.0 (99.9999\% purity) to grade 4.6 (99.996\% purity). It is interesting to note that these %%@
changes did not have any significant effect on the count rates or ratios. In previous studies at the IGISOL it had been noticed that grade %%@
6.0 helium and/or use of the Saes MonoTorr purifier have seldom affected the performance of the device. A similar remark has been recently %%@
made in the studies of the extraction efficiency and extraction time of the SHIPTRAP gas-filled stopping cell \cite{28eliseev}. In that %%@
work ordinary helium 4.6 was used without any significant degradation of the gas cell performance compared to helium 6.0.
\par
\begin{figure}
\centering
\includegraphics[width=.95\linewidth]{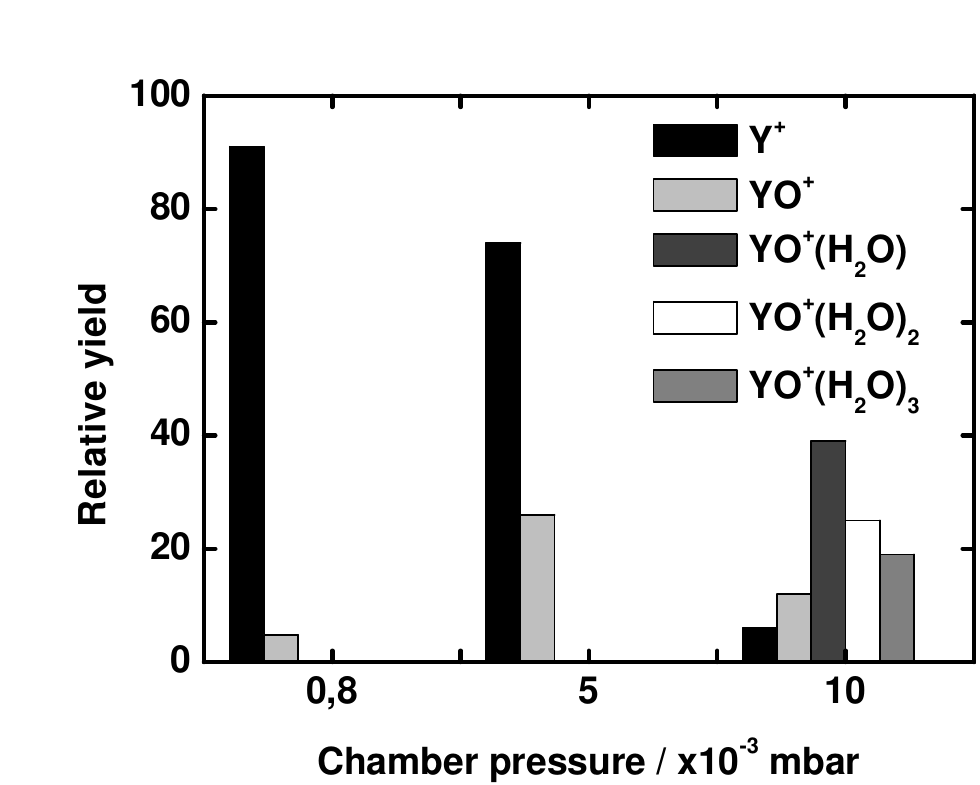}
\caption{ Relative yields of Y$^+$ and related molecules as a function of IGISOL chamber pressure. The measured pressures were $8.6 \cdot %%@
10^{-4}$, $5 \cdot 10^{-3}$ and $1 \cdot 10^{-2}$ mbar respectively.}
\label{fig:yieldcompare}
\end{figure}

A leak was introduced into the vacuum chamber and the count rates were monitored as a function of chamber pressure. Without a leak the %%@
chamber pressure was measured to be $8 \cdot 10^{-4}$ mbar with an operating ion guide pressure of 150 mbar. The chamber pressure was %%@
steadily increased until at a level of $1 \cdot 10^{-2}$ mbar a complete redistribution in the mass separated yields was identified. At %%@
higher pressures, measurements could not be reliably made as discharge within the SPIG started to occur. The distribution of molecules as a %%@
function of chamber pressure is illustrated as a histogram chart in Fig. \ref{fig:yieldcompare}. 
% Fig.21. 
The bin width of the histogram has no meaning and this figure is simply used to illustrate the relative changes in the yields. The specific %%@
pressures measured are listed in the figure caption. It should be noted that these values do not indicate the pressure in the immediate %%@
vicinity of the gas jet, and therefore should be taken as relative measurements.
\par
\begin{figure}
\centering
\includegraphics[width=.95\linewidth]{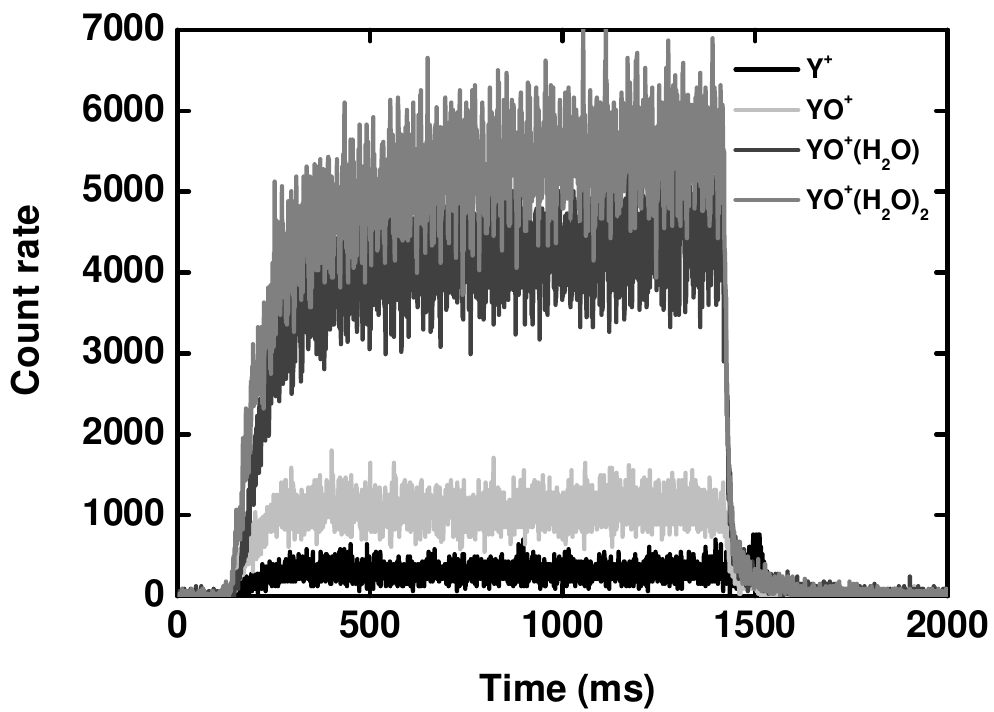}
\caption{ Evacuation time profiles of Y$^+$ and subsequent molecules detected after a leak increased the vacuum chamber pressure to $1\cdot %%@
10^{-2}$ mbar while the ion guide pressure was maintained at 150 mbar.
}
\label{fig:tofprof072}
\end{figure}

A fourth hydrate addition was also found to have a laser-related effect, however the masses above YO$^+$(H$_2$O)$_2$ had very poor mass %%@
resolution which indicated that an increase in the chamber pressure effects the pressure in the acceleration region of the SPIG. At a %%@
chamber pressure of  $1 \cdot 10^{-2}$ mbar evacuation time profiles were measured and are illustrated in Fig. \ref{fig:tofprof072}.
% Fig.22. 
 Only the lighter of the two hydrates are shown as these still had reasonable mass resolution. Yttrium is no longer the dominant ionic %%@
species detected and the yield has been considerably reduced. The most dominant molecules are now those of the two lightest hydrates, %%@
YO$^+$(H$_2$O) and YO$^+$(H$_2$O)$_2$. Exponential growth fits to the data in Fig. \ref{fig:tofprof072} yield the values listed in table %%@
\ref{tab:tofprof072}.
% Table 5
\begin{table}
\centering
\caption{ Exponential growth times in ms extracted from the data in Fig. \ref{fig:tofprof072}.}
\begin{tabular}{cccc}
\hline
$\tau_1=1/k_1\left[M\right]$   &	
$\tau_2=1/k_2\left[M'\right]$  &
$\tau_3=1/k_3\left[M'\right]$  &
$\tau_4=1/k_4\left[M'\right]$    \\ \hline
55(5)	&	55(5)	&    106(4)	&   97(7) \\ \hline
\end{tabular}
\label{tab:tofprof072}
\end{table}
\par  
	The trend of the saturation time for yttrium and yttrium oxide can be understood as the concentration of impurities is increasing, %%@
therefore via Eq.  \eqref{eq:7}, $\tau$ is expected to decrease. Although the yttrium signal is now suppressed to a level less than 10\% of %%@
the sum of all molecules, the time to reach saturation is still a factor of 10 larger than in the March 2006 data. One could imagine that %%@
close to the exit nozzle of the ion guide, before the leak is introduced, the dominant species is yttrium. This would have a time %%@
distribution profile effectively of the evacuation from the ion guide ($\approx$ 100 ms as seen in Fig. \ref{fig:tofprof07}). With a poor %%@
vacuum chamber pressure the evacuated ions may then be converted into molecules at the exit nozzle of the guide, or within the gas jet as %%@
the ions enter and pass through the SPIG. A transportation time of $\approx$ 200 $\mu$s through the SPIG for $A=100$ ions has been %%@
estimated from simulations, using typical DC voltages on the SPIG and extractor electrode, and assuming a pressure within the ion guide of %%@
150 mbar. In this simulation the pressure within the SPIG was estimated based on the modeling of the gas flow (see \cite{29pasi} for %%@
details).  Two of the more sensitive parameters to the time of flight appear to be the background pressure within the SPIG and the buffer %%@
gas velocity. At present experiments are being planned to demonstrate these effects directly. Although this transportation time is fast %%@
compared to the saturation time scales of table \ref{tab:tofprof072}, if the molecular impurity level in the SPIG region is high then %%@
molecular formation can rapidly occur via collisions within the gas jet. 
\par
\begin{figure}
\centering
\includegraphics[width=.95\linewidth]{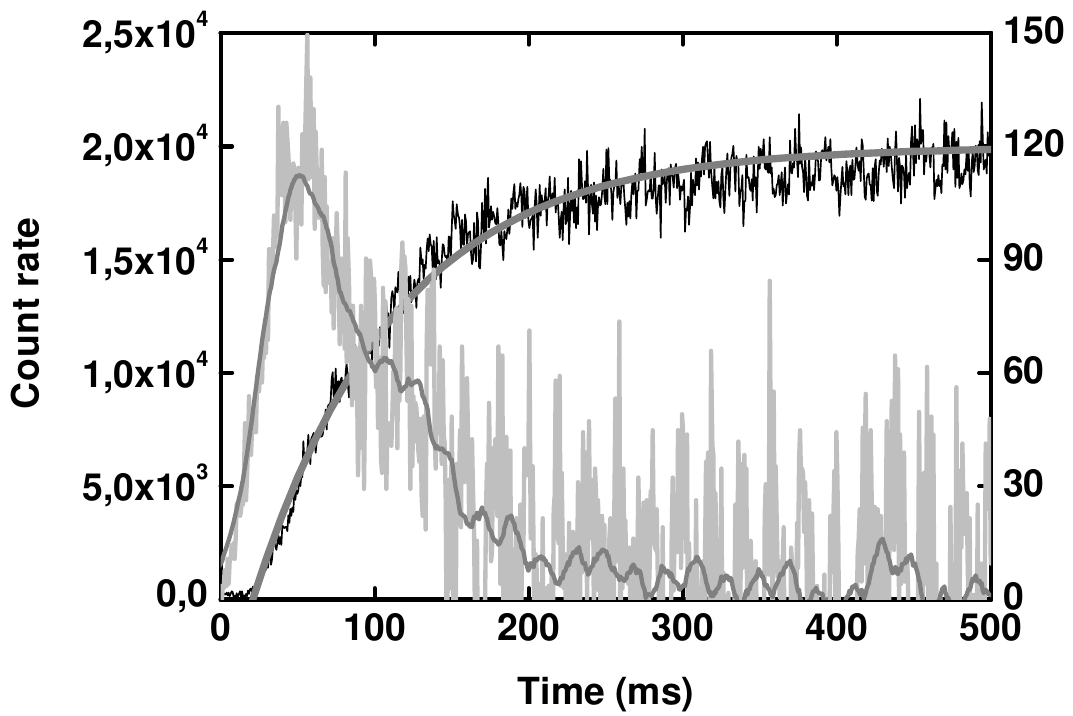}
\caption{ Summed data from Fig. \ref{fig:tofprof072} and the corresponding differentiated single shot evacuation time profile. An %%@
exponential fit to the summed profile is shown yielding a rise time of 99(4)ms and an adjacent averaging technique using 50 bins has been %%@
applied to the differentiated data in order to smooth out the fluctuations.
}
\label{fig:sumdata072}
\end{figure}

	In order to determine whether the single shot model is still valid under these conditions Fig. \ref{fig:sumdata072} shows the summation %%@
of the molecular components of Fig. \ref{fig:tofprof072}, and the subsequent single shot derivative. 
% Fig.23. 
	An exponential fit to the summed ``leak data'' gives a rise time value of 99(4) ms. Recall that the rise time of the summed data %%@
without the leak (Fig. \ref{fig:sumdata07}) yielded a value of 99(10) ms.  This is rather interesting as these two values are exactly the %%@
same, despite the fact that the summed data with no leak in the chamber is dominated by yttrium with a rise time of 96(10) ms and under %%@
poor vacuum conditions not only is the yttrium rise time reduced to 55(5) ms but it is no longer the dominant species. By worsening the %%@
vacuum conditions in the immediate chamber surrounding the ion guide the increase in impurity level reduces not only the saturation time of %%@
the yttrium time distribution profile, but also radically changes the distribution of the intensities of yttrium and subsequent molecules. %%@
What is important to note, however, is that the sum profiles which give an indication of the total yield prior to mass separation are %%@
identical, and they reflect the evacuation time profile within the ion guide. It appears that ions are not ``lost'' after the introduction %%@
of a leak into the vacuum chamber rather they are simply redistributed into other molecular forms after leaving the exit nozzle of the gas %%@
cell.
\par
\begin{figure}
\centering
\includegraphics[width=.95\linewidth]{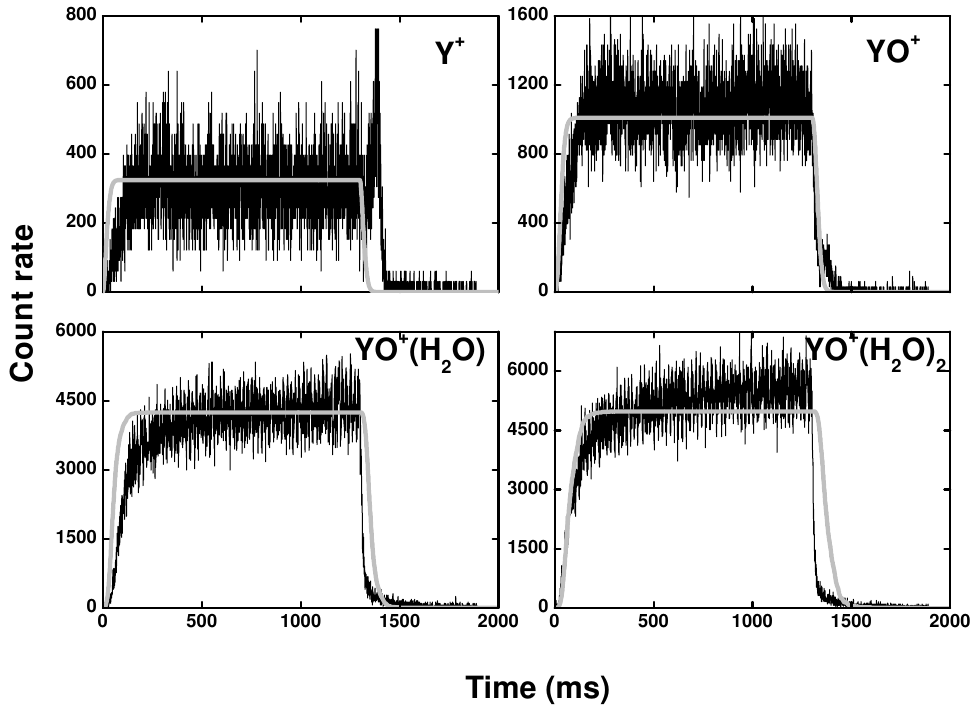}
\caption{ Comparison between the simulated single shot profiles (grey line) and the experimental data (black line).}
\label{fig:sim072}
\end{figure}

	After applying the rate equations of Eq. \eqref{eq:17} for molecular formation, the single shot evacuation time profiles are %%@
illustrated in Fig. 24 for the data taken with poor vacuum pressure in the IGISOL chamber. Recall that in Fig. \ref{fig:sim07} the rising %%@
edge and amplitude were well-reproduced however the falling edge was over-estimated. As in section \ref{sec:06} the relative error in the %%@
fitting procedure has been estimated to be $\approx$ 10\%. The time constants for molecular formation extracted from the reaction rate %%@
parameters that provide the optimum fits to reproduce the correct ratios in Fig. \ref{fig:sim072} are given in table \ref{tab:sim072}.
% Fig.24. 
% Table 6
\begin{table}
\centering
\caption{ The molecular formation time (ms) extracted from the fitted reaction rates after optimization of the ratios of the molecules.}
\label{tab:sim072}
\begin{tabular}{ccc}
\hline
$\tau_1=1/k_1\left[M\right]$   &	
$\tau_2=1/k_2\left[M'\right]$  &
$\tau_3=1/k_3\left[M'\right]$ \\ \hline
10	& 13	& 35 \\ \hline
\end{tabular}
\end{table}
	 Although the simulated ratios of the molecules appear to be in a reasonable agreement with the data, the rise times of the %%@
re-integrated single shot profiles are consistently too fast. Due to the saturation of all experimental profiles the simulated time %%@
constants can be compared with the rise times directly fitted to the data (table \ref{tab:tofprof072}). The single shot model suggests that %%@
the level of impurities within the gas cell is higher than that reflected by the experimental data. This is another indication that by %%@
introducing a leak into the chamber there are molecular processes occurring after the nozzle on fast time scales that do not affect the %%@
overall evacuation time profile extracted from the gas cell.
\par
\begin{figure}
\centering
\includegraphics[width=.95\linewidth]{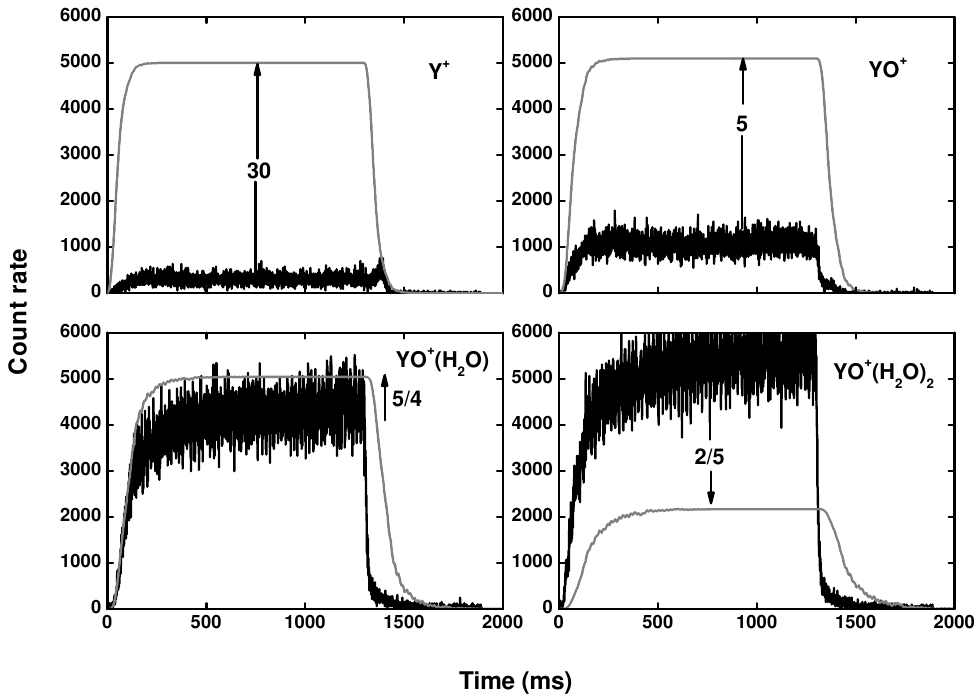}
\caption{ The experimental data (black line) of Fig. \ref{fig:sim072} fitted using the single shot model (grey line) with reaction rate %%@
parameters fixed to the time constants of table 5.}
\label{fig:sim073}
\end{figure}

	More supporting evidence for the suggestion that the presence of impurities in the environment outside the nozzle area only shifts the %%@
mass distribution of yttrium towards heavier molecules yet does not affect the total signal current or the general shape of the evacuation %%@
profile is given in Fig. \ref{fig:sim073}.
% fig 25
	 In this figure the free reaction rate parameters have been fixed to the experimental rise times given in table \ref{tab:tofprof072}. %%@
The increase of the time constants compared to those extracted from fitting the optimum molecular ratios listed in table \ref{tab:sim072}, %%@
lead to an increase in the expected rate of Y$^+$ and YO$^+$ and a decrease for the higher hydrated compounds. This is unsurprising because %%@
a longer time constant for the molecular formation infers the situation in which the resultant molecular rates should be lower for the same %%@
evacuation time of the gas cell. It should be noted that the sum of the simulation profiles appears to be very much larger than the sum of %%@
the experimental profiles. This is because there are higher mass hydrate molecules in the data (and detected with laser effects) however as %%@
they had no mass resolution it was decided to only discuss the lower mass molecules.
\par
 Finally, we note that that the higher mass hydrate %%@
additions to YO$^+$ in the 2006 data had a reasonable mass resolution. Although the baseline pressure was found to be of the same order of %%@
magnitude as that used in the measurements of Fig. \ref{fig:tofprof072}, the leak in the gas line appears to have been the dominant effect %%@
on the impurity level. This conclusion is supported not only by the good mass resolution of the molecules, but also by the clear difference %%@
in the time distribution of the hydrate additions (Fig. \ref{fig:tofprof06}).
	\par
	Interestingly, very little information is available about the importance of base pressure of the vacuum chamber to the operation of an %%@
ion guide / gas catcher system. The experimental work shown here, in combination with the use of our single shot model has implications for %%@
all gas catchers and discussions of the required impurity levels of the gas. The gas itself can be purified to a level of less than a ppb %%@
as shown in this work, however if the baseline pressure of the vacuum chamber surrounding the ion guide / gas catcher device is poor then a %%@
significant redistribution of the atomic ions into heavier molecular species on fast time scales can result. In fact, as highlighted in %%@
Fig. \ref{fig:yieldcompare}, even at a baseline vacuum pressure of $5 \cdot 10^{-3}$ mbar (measured at 150 mbar ion guide pressure) one %%@
starts to redistribute the ions.
\section{Summary and conclusions} % 4.
	The purpose of this work was to initiate a programme of research designed to lead to the successful implementation of a laser ion %%@
source for the efficient and selective production of exotic, short-lived refractory nuclei that are uniquely available at the IGISOL %%@
facility. The capabilities and flexibility of having a twin laser system running in parallel has been demonstrated with the development of %%@
a new laser ionization scheme for yttrium, in which both a dye laser and Ti:Sapphire laser were used for a second step transition providing %%@
the optimal choice of ionization scheme. In this work the laser ionization efficiency has not been explicitly discussed and is rather %%@
difficult to measure in off-line conditions. This will be addressed in a follow-up paper which discusses the role of beam ionization on the %%@
evacuation time profiles of yttrium and related molecules \cite{26moore}.
\par
	An ion guide was borrowed from the LISOL group at Louvain-la-Neuve in order to make the first off-line measurements of laser produced %%@
yttrium ions from a heated filament. A detailed study was performed to understand the effects of gas purity on an element that exhibits %%@
strong molecular formation in the presence of impurities within the gas. Results obtained from the data taken in March 2006, illustrated in %%@
the form of time-of-flight profiles, indicated that the resultant impurity level was some two orders of magnitude worse than expected from %%@
similar studies discussed in \cite{23yuri}. This highlighted the importance of careful control of the gas-handling system. In order to gain %%@
a more detailed understanding of the competition between the molecular formation process and evacuation from the ion guide, the %%@
experimental data was analyzed in the framework of a series of molecular rate equations in order to extract time profiles of single laser %%@
shots.
	\par
A repeat of the experiment was performed in March 2007 in which the lasers ionized the yttrium atoms for a time period such that a clear %%@
saturation level could be achieved. With a better control over the gas purification a clear reduction in the level of impurities to sub-ppb %%@
could be achieved. The importance of the vacuum pressure in the immediate vicinity of the ion guide was apparent after a controlled leak %%@
was added directly to the vacuum chamber. What is rather astonishing is that although the increase in the number of impurities reduces the %%@
time needed to reach a saturation level in yttrium, by summing the individual yttrium and associated molecular time-of-flight profiles the %%@
total evacuation profile with and without the added leak has the same overall shape. This time profile reflects the evacuation time of the %%@
laser ions from within the ion guide. It appears that ions are not ``lost'' after the introduction of a leak into the vacuum chamber rather %%@
they are simply redistributed into other molecular forms after leaving the exit nozzle of the gas cell.
\par
The development of the single shot model provides a new tool with which to extract from the data direct information of the level of %%@
impurities within the gas cell. This removes the need for complicated gas flow simulations which would be required if the evacuation time %%@
of the ions within the laser path was to be accurately described and then deconvoluted from the experimental time-of-flight profile in %%@
order to separate the effect due to chemistry. The issue of where the molecular formation is happening has been successfully addressed in %%@
this work and has led to the need for a better knowledge of the time of flight of ions through the sextupole ion guide, and the parameters %%@
which can most influence this time scale. In parallel, simulations are now underway both in Jyväskylä and at the NSCL facility, Michigan %%@
State University, in order to study the drift time of ions through the SPIG used in this work. This effort reflects not only the interest %%@
but concern of the gas catcher community as long drift times can result in losses of the ion of interest through molecular formation and %%@
possibly charge exchange.
\par
In studies performed at Louvain-la-Neuve with a cobalt filament and a 30 MeV cyclotron proton beam passing through a gas cell filled with %%@
500 mbar argon, the final conclusion was that in order to explain the resulting molecular sidebands either the impurity of the buffer gas %%@
was higher than 1 ppb, or the molecular ions are formed close to the exit hole of the gas cell or in the gas jet leaving the ion source %%@
\cite{30facina}. The present work provides a unique method with which to separate these two scenarios in the development of the single shot %%@
model. The need for extreme gas purity is common to all existing and planned gas cell devices. Much effort has been put into cleaning up %%@
the gas down to the sub-ppb impurity level, however even at this low level the molecular formation still seems to be a problem in some %%@
cases. Partly because of these difficulties, cryogenic gas catchers are currently being developed \cite{31dendooven}. 
In these devices it is hoped that the impurities can be frozen out by cooling the gas and the gas catcher / ion guide system to liquid %%@
nitrogen temperature. Future plans exist at the IGISOL facility to combine the new laser ion source with a cryogenic ion guide. In this %%@
manner, similar to the present work, the evolution of the molecular sidebands on chemically reactive elements such as yttrium can be %%@
monitored as a function of the cooling temperature. It will be important though not only to develop and incorporate these cryogenic %%@
techniques further, but to ensure a clean environment through which the ions must pass once they are extracted from the gas cell.
\section*{Acknowledgements}
This work has been supported by the LASER Joint Research Activity project under the EU 6th Framework program ``Integrating Infrastructure %%@
Initiative - Transnational Access'', Contract number: 506065 (EURONS) and by the Academy of Finland under the Finnish centre of Excellence %%@
Program 2006-2011 (Nuclear and Accelerator Based Physics Program at JYFL).

\bibliographystyle{elsart-num}
\bibliography{yttrium_offline}

\end{document}